\documentclass[11pt]{article}
\pdfoutput=1
\usepackage{jheppub,amsmath,amssymb}
\usepackage{enumerate}
\usepackage{bm}
\usepackage{bbm}
\usepackage[usenames,dvipsnames]{xcolor}
\usepackage{graphics}
\usepackage{graphicx}
\usepackage{tikz,todonotes}
\usepackage[utf8]{inputenc}
\usetikzlibrary{arrows,positioning,decorations.markings,decorations.pathmorphing,calc}

\definecolor{hyperref}{RGB}{026,028,087}

\def\gsim{ \lower .75ex \hbox{$\sim$} \llap{\raise .27ex \hbox{$>$}} }
\def\lsim{ \lower .75ex \hbox{$\sim$} \llap{\raise .27ex \hbox{$<$}} }
\def\be{\begin{equation}}
\def\ee{\end{equation}}
\def\bea{\begin{eqnarray}}
\def\eea{\end{eqnarray}}
\newcommand{\nn}{\nonumber}

\def \bal#1\eal  {\begin{align} #1 \end{align}}

\newcommand{\pd} {\partial}

\newcommand{\ba}{\begin{array}}
\newcommand{\ea}{\end{array}}

\newcommand{\commentout}[1]{}

\newcommand{\comment}[1]{}

\newcommand{\bs}{\begin{split}}

\def\ba{\begin{eqnarray}}
\def\ea{\end{eqnarray}}

\def\nn{\nonumber}

\def\d{\mathrm{d}}

\def\mupn{^\mu_{\, \nu}}

\def\({\left(}
\def\){\right)}

\newcommand*{\mathcolor}{}
\def\mathcolor#1#{\mathcoloraux{#1}}
\newcommand*{\mathcoloraux}[3]{%
  \protect\leavevmode
  \begingroup
    \color#1{#2}#3%
  \endgroup
}
\newlength{\stheight}
\newcommand\textst[1][fu-grey]{
	\ifmmode\setlength{\stheight}{+1.0ex}
	\else\setlength{\stheight}{+0.5ex}
	\fi
	\bgroup\markoverwith{\textcolor{#1}{\rule[\the\stheight]{2pt}{1.0pt}}}\ULon
}
\newcommand{\textins}[2][fu-grey]{
	\ifmmode\mathcolor{#1}{#2}
	\else\textcolor{#1}{#2}\@\,
	\fi
}

\def\({\left(}
\def\){\right)}

\allowdisplaybreaks[3]

\def\ba{\begin{eqnarray}}
\def\ea{\end{eqnarray}}

\def\stu{St\"uckelberg }

\def\d{\mathrm{d}}

\def\mupn{^\mu_{\, \nu}}
\def\({\left(}
\def\){\right)}

\def\mpl{M_{\rm Pl}}

\begin{document}

\title{Improved Positivity Bounds and Massive Gravity}

\author[a,b]{Claudia de Rham}
\author[a]{Scott Melville}
\author[a,b]{Andrew J. Tolley}
\affiliation[a]{Theoretical Physics, Blackett Laboratory, Imperial College, London, SW7 2AZ, U.K.}
\affiliation[b]{CERCA, Department of Physics, Case Western Reserve University, 10900 Euclid Ave, Cleveland, OH 44106, USA}

\emailAdd{c.de-rham@imperial.ac.uk}
\emailAdd{s.melville16@imperial.ac.uk}
\emailAdd{a.tolley@imperial.ac.uk}

\abstract{
Theories such as massive Galileons and massive gravity can satisfy the presently known improved positivity bounds provided they are weakly coupled. We discuss the form of the EFT Lagrangian for a weakly coupled UV completion of massive gravity which closely parallels the massive Galileon, and perform the power counting of corrections to the scattering amplitude and the positivity bounds. The Vainshtein mechanism which is central to the phenomenological viability of massive gravity is entirely consistent with weak coupling since it is classical in nature. We highlight that the only implication of the improved positivity constraints is that the EFT cutoff is lower than previous assumed, and discuss the observable implications, emphasizing that these bounds are not capable of ruling out the model contrary to previous statements in the literature.}

\maketitle


\section{Introduction}

In the last decade a number of infrared modified theories of gravity have been developed to provide a new perspective on dark energy and the cosmological constant problem. Lorentz invariant massive gravity theories are perhaps the simplest and physically best motivated examples of these theories, where we imagine that some hidden `dark energy/gravity' sector spontaneously breaks diffeomorphisms, giving a mass to the graviton. An explicit UV theory capable of doing this is not known at present, but it is known how to write down the effective field theory (EFT) for the Goldstones/\stu fields of the spontaneously broken symmetry \cite{ArkaniHamed:2002sp}. A generic such theory is known to break perturbative unitarity at the exceptionally low scale $\Lambda_5 = (m^4 \mpl)^{1/5}$, where $m$ is the graviton mass, which is too low to make much use of these theories phenomenologically. It is, however, possible to engineer the form of the mass term so that the scale of perturbative unitarity violation is raised to the scale $\Lambda_3 = (m^2 \mpl)^{1/3}$, significantly improving the observational window of usefulness of these EFTs \cite{deRham:2010ik,deRham:2010kj,deRham:2014zqa}. Remarkably, the $\Lambda_3$ theories of massive gravity are classically free of ghosts, meaning that they propagate nonlinearly 5 degrees of freedom \cite{deRham:2010kj,Hassan:2011hr}.
\\

From the very beginning of the development of these theories, it was clear that they exhibited a decoupling or double scaling limit, obtained by sending $m \rightarrow 0$ and $\mpl \rightarrow \infty$ while keeping $\Lambda_3$ fixed, and that in this limit the dynamics of the helicity-0 and helicity-1 modes is equivalent to a Galileon theory \cite{Nicolis:2008in} coupled to a Maxwell vector \cite{deRham:2009rm,deRham:2010gu,deRham:2010ik,deRham:2010kj,Tasinato:2012ze,Ondo:2013wka,Gabadadze:2013ria}. This simple fact creates an immediate tension in the possible UV completion of these theories since it has been argued by means of S-matrix analyticity (positivity bounds) that the massless Galileon does not admit a standard, local, Lorentz invariant UV completion \cite{Adams:2006sv}.  There have been three possible solutions to this:

\begin{enumerate}[{\bf Approach} 1]

\item Theories of massive gravity, like the massless Galileon, admit a nonstandard UV completion, where some of the standard requirements are discarded (e.g. strict locality).
    \label{appro1}

\item Theories of massive gravity should be coupled to new light states which play a crucial role in recovering consistency of the positivity bounds.
    \label{appro2}

\item Theories of massive gravity may admit a standard Wilsonian UV completion in which the terms arising beyond the decoupling limit can be used to respect the positivity bounds.
    \label{appro3}

\end{enumerate}

For example in approach \ref{appro1}, it is possible to give up the requirement of polynomial boundedness of the S-matrix (locality) without discarding analyticity, and this changes the import of the positivity bounds since the Froissart bound need no longer apply \cite{Keltner:2015xda}. Indeed, it is far from clear that a gravitational theory should exhibit locality in the same sense as a local field theory \cite{Giddings:2001pt,Giddings:2006vu}, and it is expected that polynomial boundedness is violated for massless graviton scattering amplitudes at least away from the forward scattering limit \cite{Giddings:2009gj}.
In this approach, the strong coupling scale $\Lambda_3$ may not necessarily imply the existence of new states, as in a standard Wilsonian completion, but may exhibit some more fundamentally non-perturbative UV completion \cite{Dvali:2010jz,Dvali:2010ns,Dvali:2011nj,Dvali:2011th}, and the failure of the usual positivity constraints may be a signature of this \cite{Dvali:2012zc}. \\

An example of approach \ref{appro2} is the `warped massive gravity' model considered in \cite{Gabadadze:2017jom}, which is a five-dimensional brane model that describes in four dimensions an effectively massive gravity theory but with a non-local mass term due to a continuum of light states. These light states modify the usual assumptions, of a single pole at mass $m$ and a branch cut beginning at $4m^2$, which are used in the usual derivations of the positivity bounds. \\

In what follows, we shall focus exclusively on approach \ref{appro3}, which is the most conservative\footnote{It is important to emphasize that whenever conclusions are drawn as to the possible existence of a UV completion with with our assumption of being in the third approach, this does not preclude the existence of a UV completion along the lines of the first or second approaches. }. The basic tension of this approach is that the improved positivity bounds we will discuss later imply that:
\ba
\(\begin{array}{c}
\text{\bf Terms which vanish as }\\
m \rightarrow 0
\end{array} \)>
\(\begin{array}{c}
\text{\bf Terms which are finite as }\\
m \rightarrow 0 {\text{ \bf but vanish when }} g_* \rightarrow 0
\end{array} \)\nn\,,
\ea
where $g_*$ is a coupling constant effectively playing the role of $\hbar$. Given the phenomenologically desired small mass for the graviton, the third approach then necessarily forces us into considering weakly coupled $g_* \ll 1$ theories\footnote{In this manuscript, the value of $g_*$ determines whether or not we are in `weak/strong coupling'. This differs from the notion of `strong coupling' sometimes referred to, which is when perturbative unitarity breaks down for a given irrelevant operator along the lines of \cite{deRham:2014wfa}.}. In an interesting recent study, Ref.~\cite{Bellazzini:2017fep} used the improved positivity bound to constrain massive gravity with the prior  assumption that $g_*\gg 1$, which would  of course be in tension with this approach.  In this study, we show that the improved positivity bounds are fully consistent with the Cheung \& Remmen ``island of positivity'' \cite{Cheung:2016yqr} when $g_* \ll 1$, which is entirely compatible with the Vainshtein mechanism \cite{Vainshtein:1972sx}.

We begin by reviewing the origin of the improved positivity bounds, and their application to a weakly coupled massive Galileon \cite{deRham:2017imi} in section~\ref{massivegalileon}, before moving onto the closely parallel discussion for massive gravity in section~\ref{massivegravity}, and close with a discussion of the Vainshtein region in weakly coupled massive gravity in section~\ref{sec:EnteringVain}.

\section{Origin of Improved Positivity Bounds}
\label{sec:ImpPositivity}

The original positivity bounds which applied to the forward scattering limit were derived for scalars in \cite{Adams:2006sv}. In the forward limit, these have been extended to particles of all spins in \cite{Bellazzini:2016xrt} and for example utilized to put constraints on `approach \ref{appro3}' UV completions of massive gravity\footnote{Other S-matrix constraints on massive spin-2 fields have been considered in \cite{Hinterbichler:2017qyt} and complement these (improved) positivity bounds. } in \cite{Cheung:2016yqr,Bonifacio:2016wcb}. These were extended for scalars beyond the forward scattering limit in \cite{deRham:2017avq} (for earlier work see \cite{Pennington:1994kc,Vecchi:2007na,Manohar:2008tc,Nicolis:2009qm,Bellazzini:2014waa}) and for general spin particles in \cite{deRham:2017zjm}.

\subsection{Positivity bounds}

For example, in the case of spin-zero scattering the general positivity bounds applied to a $2 \rightarrow 2$ scattering amplitude $A(s,t)$ take the form \cite{deRham:2017avq}
\be
\label{generalbounds}
Y^{(2N,M)} (t) >0 ~~~ \text{for}~~~  N \ge 1 \, , \, M \ge 0  \, ,\quad 0  \le t < 4m^2  \,,
\ee
where $Y^{(2N,M)}(t)$ is defined by the following recurrence relation
\ba
 && \hspace{-15pt} Y^{(2N,0)} (t)  = B^{(2N,0)}(t)\,,\\
&& \hspace{-15pt}  Y^{(2N,M)} (t) = \sum_{r=0}^{M/2} c_r B^{(2(N+r),M-2r)}  + \frac{1}{{\cal M}^2} \sum_{ \text{even}\,k=0}^{(M-1)/2}  (2(N+k)+1) \beta_k   Y^{(2(N+k),M-2k-1)},\quad
\label{eqn:Y}
\ea
where ${\cal M}^2 = {\rm Min}( \mu + t /2-2m^2)=2m^2+t/2$,
and the coefficients $c_r$ and $\beta_k$ are defined recursively by,
\ba
\label{eq:coeffs}
c_k=-\sum_{r=0}^{k-1}\frac{2^{2(r-k)}c_r}{(2(k-r))!}\,, \quad {\rm with }\quad c_0=1\,,
\quad
{\rm and }
\quad
\beta_k=(-1)^k \sum_{r=0}^k \frac{2^{2(r-k)-1}}{(2(k-r)+1)!}c_r \ge 0\,.\qquad
\ea
Here $B^{(N,M)}$ denotes
\be
B^{(N,M)}(t)  = \frac{1}{M!}
 \left. \pd^N_v \pd^M_t   B(v,t)\right|_{v=0}\,,
\ee
where $B(v,t)$ is the pole-subtracted scattering amplitude\footnote{It is not strictly necessary to subtract the $t$-channel pole, but for spin-zero scattering is more elegant.},
\ba
B(v,t)  = A(s,t)-\frac{\lambda}{m^2-s} -\frac{\lambda}{m^2-t} -\frac{\lambda}{m^2-u} \,,
\ea
with  $v=s+t/2-2m^2$, and analyticity together with the Froissart bound implies that
\ba
 B(v,t)=b(t)+\int_{4m^2}^\infty \frac{2 \d \mu }{\pi}\frac{v^2}{(\mu+t/2-2m^2)}
\frac{{\rm Im}A(\mu,t)}{(\mu+t/2-2m^2)^2-v^2}\,.
\ea
The generalization of these bounds to elastic scattering for two particles of arbitrary spin is given in \cite{deRham:2017zjm}.

\subsection{Improved Positivity Bounds}

As pointed out in \cite{deRham:2017imi} (see also \cite{Bellazzini:2016xrt}) these basic positivity bounds can be improved by using knowledge of the right-hand side of the inequalities, evaluated in the regime in which perturbation theory is valid. Specifically, assuming that perturbation theory can be trusted up to a scale $\epsilon \Lambda \gg m $ where $\epsilon  \ll 1$, then we may define (see \cite{deRham:2017avq,deRham:2017imi})
\be
B_{\epsilon \Lambda}^{(2N,M)}(t) = B^{(2N,M)}(t) - \sum_{k=0}^M \frac{2 (-1)^k}{\pi k! 2^k} \frac{(2N+k)!}{(M-k)!} \int_{4m^2}^{\epsilon^2 \Lambda^2} \d \mu \frac{\partial_t^{M-k} {\rm Im} \, A(\mu,t)}{(\mu+t/2-2m^2)^{2N+k+1}} \, .
\ee
We can then compute $Y_{\epsilon \Lambda}^{(2N,M)}(t)$ out of $ B_{\epsilon \Lambda}^{(2N,M)}(t) $ via the recurrence relations defined in (\ref{eqn:Y}) where we now take ${\cal M}^2 =  \epsilon^2 \Lambda^2 + t /2-2m^2 \approx \epsilon^2 \Lambda^2$. Following the arguments of \cite{deRham:2017avq} it follows that
\be
\label{eq:improvedbounds}
Y_{\epsilon \Lambda}^{(2N,M)}(t) >0 \, .
\ee
For example, focusing on the case of no $t$ derivatives, $M=0$, then we have
\be
Y_{\epsilon \Lambda}^{(2N,0)}(t) = B^{(2N,0)}-\frac{2 }{\pi} (2N)! \int_{4m^2}^{\epsilon^2 \Lambda^2} \d \mu \frac{ {\rm Im} \, A(\mu,t)}{(\mu+t/2-2m^2)^{2N+1}} >0 \, ,\quad N \ge 1\,,
\ee
and further focusing on $N=1$ we have
\be
\left. B^{(2,0)}(t)=\pd_v^2   B(v,t) \right|_{v=0} > \frac{4 }{\pi}  \int_{4m^2}^{\epsilon^2 \Lambda^2} \d \mu \frac{ {\rm Im} \, A(\mu,t)}{(\mu+t/2-2m^2)^{3}} \, .
\ee
Further focusing on the forward scattering limit, $t=0$, and making use of the optical theorem to relate the imaginary part of the scattering amplitude to the total cross section $\sigma_{\rm total}(s)$ for the two particles to scatter, then we have
\be
\label{bound1}
 B^{(2,0)}(0)> \frac{4 }{\pi}  \int_{4m^2}^{\epsilon^2 \Lambda^2} \d \mu \frac{ {\rm Im} \, A(\mu,0)}{(\mu-2m^2)^{3}} = \frac{4 }{\pi}  \int_{4m^2}^{\epsilon^2 \Lambda^2} \d \mu \, \sqrt{1- \frac{4 m^2}{\mu^2}}\frac{\mu \sigma_{\rm total}(\mu) }{(\mu-2m^2)^{3}}  \, .
\ee
Since we are now in the forward scattering limit, this bound is easily generalized to elastic scattering of two arbitrary helicity particles $\lambda_1$ and $\lambda_2$, and we obtain\footnote{The only difference being that for $ B_{\lambda_1 \lambda_2}(v,t)$ we do not subtract the $t$-channel pole since its residue is itself $s$ dependent \cite{deRham:2017zjm}.}
\be
\label{bound2}
 B_{\lambda_1 \lambda_2}^{(2,0)}(0) >  \frac{2 }{\pi}  \int_{4m^2}^{\epsilon^2 \Lambda^2} \d \mu \, \sqrt{1- \frac{4 m^2}{\mu^2}} \left[  \frac{\mu \sigma^{\lambda_1 \lambda_2}_{\rm total}(\mu) }{(\mu-2m^2)^{3}}+\frac{\mu \sigma^{\lambda_1 - \bar{\lambda}_2}_{\rm total}(\mu) }{(\mu-2m^2)^{3}} \right]  \, ,
\ee
where the combination $\sigma^{\lambda_1 - \bar{\lambda}_2}_{\rm total}(\mu)$ arises via crossing symmetry, since in the forward limit crossing flips the sign of the helicity. This is a result recently considered in \cite{Bellazzini:2017fep}. We see that this is just a special case of the general procedure for improving all such positivity bounds given in (\ref{eq:improvedbounds}) \cite{deRham:2017avq} and generalized to all spins in \cite{deRham:2017zjm}. The next order forward limit bound is for example
\be
 B_{\lambda_1 \lambda_2}^{(4,0)}(0) >  \frac{4! }{\pi}  \int_{4m^2}^{\epsilon^2 \Lambda^2} \d \mu \, \sqrt{1- \frac{4 m^2}{\mu^2}} \left[  \frac{\mu \sigma^{\lambda_1 \lambda_2}_{\rm total}(\mu) }{(\mu-2m^2)^{5}}+\frac{\mu \sigma^{\lambda_1 - \bar{\lambda}_2}_{\rm total}(\mu) }{(\mu-2m^2)^{5}} \right]  \, .
\ee
\subsection{Mixing of orders}
\label{Mixing}

The improved positivity bounds considered in \cite{Bellazzini:2016xrt,deRham:2017imi} and more recently in \cite{Bellazzini:2017fep} have the feature that they mix orders of tree and loops. From a purely perturbative thinking, it seems peculiar to place a constraint on the tree level amplitude by a contribution that arises at one-loop order. The origin of the matching of tree and loop orders implied by the optical theorem applies to the imaginary part of the scattering amplitude, but it does not apply to the real part. The role of analyticity is to determine the real part from the imaginary part, and the assumption of the Froissart bounds imposes a nontrivial constraint on the real part which effectively mixes orders. \\

To see how this works, we can apply this formalism to  the example of the massive Galileon which will be considered in section \ref{massivegalileon}.
The tree level amplitude is assumed to behave as
\be
A_{\rm tree}(s,t) = g_*^2 A_0 \( \frac{s}{\Lambda^2} ,\frac{t}{\Lambda^2} \) \, ,
\ee
where as we shall see in section~\ref{massivegalileon}, the leading contribution to the tree-level amplitude for $s,t \gg m^2$ is $A_0=(s^3+t^3+u^3)/\Lambda^6+\cdots$. The $n$-loop amplitudes are of the form
\be
A_{n-\rm loop}(s,t) = g_*^{2+2n} A_n \( \frac{s}{\Lambda^2} ,\frac{t}{\Lambda^2} \) \, .
\ee
If we compute the scattering amplitude at one loop, then we know via the optical theorem that its imaginary part will be fixed in terms of its tree level contribution, so that ${\rm Im}(A_{\rm one-loop}(s,t)) \sim |A_{\rm tree}|^2$ grows as $g_*^4 s^6/\Lambda^{12}$ for $s \gg m^2$, i.e. $A_1 \sim s^6/\Lambda^{12}$. We further know that $A_{\rm one-loop}(s,t)$ is a crossing symmetric, analytic function of $s$ with the usual poles and branch cuts. It thus follows due to the growth of the imaginary part that the form of the one-loop amplitude is an analytic function with 8 subtractions\footnote{Actually only 7 are necessary but by crossing symmetry the odd terms in $v$ vanish. Similarly, this crossing symmetry implies only $4$ subtraction functions $c_n (t)$ are non-zero.}
\ba
&&B_{\rm tree}(s,t)+  B_{\rm one-loop}(s,t)  =  \sum_{n=0}^3 c_{n}(t) v^{2n}  \nn \\
&&  + \frac{1}{\pi} v^8 \int_{4m^2}^{\infty} \d \mu \frac{ {\rm Im}(A_{\rm one-loop}(\mu,t))}{(\mu-2m^2+t/2)^8 (\mu-s)}+\frac{1}{\pi} v^8 \int_{4m^2}^{\infty} \d \mu \frac{ {\rm Im}(A_{\rm one-loop}(\mu,t))}{(\mu-2m^2+t/2)^8 (\mu-u)} \, ,\qquad
\ea
with  $v=2m^2+s-t/2$. The subtraction functions, which are undetermined by analyticity, are related to the power law and log divergences which are removed by the addition of local counterterms and so have the form
\be
c_n(t) = c^{\rm tree}_n(t) + c^{\text{one-loop}} _n(t)\, .
\ee
where $c^{\rm tree}_n(t)$ comes from the leading classical Lagrangian and $ c^{\text{one-loop}}_n(t)$ are the one-loop contributions.  This split is sensitive to the renormalization prescription, but $c_n(t)$ is not.\\

The power of analyticity comes from the statement that we expect the exact amplitude to respect the Froissart bound (unlike the one-loop amplitude which explicitly violates it) which means that it is sufficient to perform only 2 subtractions. This can be rearranged as
\ba
&&  B_{\rm exact}(s,t) =a(t)+ \sum_{n=1}^3  \frac{2}{\pi} v^{2n} \int_{4m^2}^{\infty} \d \mu \frac{ {\rm Im}(A_{\rm exact}(\mu,t))}{(\mu-2m^2+t/2)^{2n+1} }\nn \\
&&  + \frac{1}{\pi} v^8 \int_{4m^2}^{\infty} \d \mu \frac{ {\rm Im}(A_{\rm exact}(\mu,t))}{(\mu-2m^2+t/2)^8 (\mu-s)}+\frac{1}{\pi} v^8 \int_{4m^2}^{\infty} \d \mu \frac{ {\rm Im}(A_{\rm exact}(\mu,t))}{(\mu-2m^2+t/2)^8 (\mu-u)} \, ,
\ea
and thus matching $A_{\rm exact}(s,t)   = A_{\rm tree}(s,t)+A_{\rm one-loop}(s,t)  + \dots$,
we infer $a(t) \approx c_0(t)$, and for the other subtraction functions
\be
c_n(t) = \frac{1}{(2n)!}\(B_{\rm tree}^{(2n,0)}(t)+B_{\rm one-loop}^{(2n,0)}(t)\)\approx \frac{2}{\pi}  \int_{4m^2}^{\infty} \d \mu \frac{ {\rm Im}(A_{\rm exact}(\mu,t))}{(\mu-2m^2+t/2)^{2n+1} } \, , \quad 1 \le n \le 3 \, .
\ee
Choosing an energy scale $E =\epsilon \Lambda \gg m$ with $\epsilon < 1$ sufficiently small that ${\rm Im}(A_{\rm exact}(E^2,t)) \approx {\rm Im}(A_{\rm one-loop}(E^2,t)) $ then
\ba
c^{\rm tree} _n(t)+ c^{\text{one-loop}} _n(t) \approx \frac{2}{\pi}  \int_{4m^2}^{E^2} \d \mu \frac{ {\rm Im}(A_{\rm one-loop}(\mu,t))}{(\mu-2m^2+t/2)^{2n+1} }+\frac{2}{\pi} \int_{E^2}^{\infty} \d \mu \frac{ {\rm Im}(A_{\rm exact}(\mu,t))}{(\mu-2m^2+t/2)^{2n+1} }\,,\qquad
\ea
where the equality is true up to two-loop corrections. Thus the tree plus one-loop renormalization coefficient $c_n(t)$ is bounded by the one-loop imaginary part,
\be
c^{\rm tree} _n(t)+ c^{\text{one-loop}} _n(t) \gtrapprox  \frac{2}{\pi}  \int_{4m^2}^{E^2} \d \mu \frac{ {\rm Im}(A_{\rm one-loop}(\mu,t))}{(\mu-2m^2+t/2)^{2n+1} } \, , \quad 1 \le n \le 3 \, .
\ee
In the case of $c_n(t)$ for the massive Galileon or massive gravity, the right-hand side can be computed using the optical theorem, and scales as
\be
g_*^4 \frac{E^{12-4n}}{\Lambda^{12}} \, , \quad 1 \le n<3 \, , \quad g_*^4 \frac{ \ln E }{\Lambda^{12}} \, , \quad n=3 \, ,
\ee
the growth in energy being a reflection of the non-renormalizability of the leading interactions, whereas the one-loop counterterm necessary to renormalize log divergences\footnote{In general in an EFT it is consistent to assume that all loops are calculated in dimensional regularization \cite{Burgess:1992gx} provided all allowed higher dimension operators are included in the tree level Lagrangian, since a different regularization scheme can be absorbed in a redefinition of the tree level interactions, here $c^{\rm tree} _n(t)$. Regardless, no regularization scheme that respects the Galileon symmetry will give terms which give an unsuppressed contribution to $c_2(t)$.} from the leading Galileon operators behaves as $c_n(t)^{\text{one-loop}}  \sim g_*^4 \frac{m^{12-4n}}{\Lambda^{12}}$. We thus find
\be
\frac{c_n(t)^{\text{one-loop}}}{\frac{2}{\pi}  \int_{4m^2}^{E^2} \d \mu \frac{ {\rm Im}(A_{\rm one-loop}(\mu,t))}{(\mu-2m^2+t/2)^{2n+1} }} \sim \( \frac{m}{E}\)^{12-4n} \ll 1  \,  , \text{ for } 1 \le n < 3\, .
\ee
In other words, the calculated one-loop contribution to the real part of the amplitude is negligible with respect to the real part inferred via analyticity from the one-loop imaginary part. For $n=3$, the two terms are comparable. Thus for $1 \le n < 3$, it has to be the tree level part of $c_n(t)$ that matches onto the latter:
\be
c^{\rm tree}_n(t)  \gtrapprox  \frac{2}{\pi} \int_{4m^2}^{E^2} \d \mu \frac{ {\rm Im}(A_{\rm one-loop}(\mu,t))}{(\mu-2m^2+t/2)^{2n+1} }   \,  , \text{ for } 1 \le n < 3\, .
\ee
This mixing of orders will be a characteristic of any system in which the low energy cross section grows faster than allowed by the Froissart bound and for which similar reasoning applies. \\

Continuing to two loops, the amplitude scales as $A_{\rm two-loop} \sim \frac{g_*^6 E^{18}}{\Lambda^{18}}$ at large energies and so it will be given by an analytic function with 10 subtractions
\ba
&&B_{\rm tree}(s,t)+  B_{\rm one-loop}(s,t) +  B_{\rm two-loop}(s,t)  = \sum_{n=0}^4 c_{n}(t) v^{2n}  \nn \\
&&  + \frac{1}{\pi} v^{10} \int_{4m^2}^{\infty} \d \mu \frac{ {\rm Im}(A_{\rm one-loop}(\mu,t)+A_{\rm two-loop}(s,t))}{(\mu-2m^2+t/2)^{10} (\mu-s)} + {s \leftrightarrow u} \, .
\ea
Comparing with the exact amplitude we again have for $1 \le n \le 4$,
\ba
c_n(t)= \frac{1}{(2n)!}B^{(2n,0)}(t) &=& \frac{2}{\pi}  \int_{4m^2}^{\infty} \d \mu \frac{ {\rm Im}(A_{\rm exact}(\mu,t))}{(\mu-2m^2+t/2)^{2n+1} }  \\
&\gtrapprox&   \int_{4m^2}^{E^2} \d \mu \frac{ {\rm Im}(A_{\rm one-loop}(\mu,t))}{(\mu-2m^2+t/2)^{2n+1} } + \int_{4m^2}^{E^2} \d \mu \frac{ {\rm Im}(A_{\rm two-loop}(\mu,t))}{(\mu-2m^2+t/2)^{2n+1} } \, . \nn
\ea
For $n=1,2,3$, the first term will dominate since $A_{\rm two-loop}/A_{\rm one-loop} \sim g_*^2 (E/\Lambda)^6 \ll 1$ and so we recover the one-loop result
\be
c_n(t)  \gtrapprox   \int_{4m^2}^{E^2} \d \mu \frac{ {\rm Im}(A_{\rm one-loop}(\mu,t))}{(\mu-2m^2+t/2)^{2n+1} } \left( 1 + {\cal O} \( g_*^2 \(\frac{E}{\Lambda} \)^6\) \right) \, ,
\ee
which will give the same constraint on $c^{\rm tree}_n(t)$ as before. For $n=4$, if $g_*^2 \ln(E/m) \gg 1$ it is the second term that dominates, since the first integral converges in the limit $E \rightarrow \infty$ whereas the second continues to grow logarithmically in $E$,
\be
c_4(t)  \gtrapprox   \int_{4m^2}^{E^2} \d \mu \frac{ {\rm Im}(A_{\rm two-loop}(\mu,t))}{(\mu-2m^2+t/2)^{9} } \left( 1 + {\cal O} \(\frac{1}{g_*^2 \ln(E/m)}\) \ \right) \, ,
\ee
however if $g_*^2 \ln(E/m) \ll 1$ the one-loop continues to dominate.  This procedure will continue at higher loops, with higher subtraction polynomials being determined principally by higher loop imaginary parts. In this sense, the loop expansion is entirely consistent, despite the mixing of orders.

\section{Weakly Coupled Massive Galileons}

\label{massivegalileon}
We now explore the implications of these bounds to massive Galileons as a warm-up to massive gravity. We emphasize again that, in doing so, we force ourselves into approach \ref{appro3}.  We thus assume the existence of a weakly coupled UV completion for the Galileon for which the low energy EFT may be structured in terms of a single scale $\Lambda$ and a small coupling constant $g_*$ in the form \cite{deRham:2017imi}
\be
{\cal L} = \frac{1}{g_*^2}\left( \Lambda^4 L_0(\pi/\Lambda , \partial/\Lambda ) + g_*^2 L_1 + \dots \right)\,,
\ee
where $L_0$ is the tree level part and $L_1$ and higher are understood to come from loops of the heavy fields integrated out.
This `single scale - single coupling' assumption has been considered in \cite{Giudice:2007fh,Liu:2016idz} and is naturally preserved under loops. The system becomes strongly coupled in the limit where  $g_*
\to 4\pi$. There are many examples in which the UV theory is weakly coupled and preserve this structure.
For example, weakly coupled string theory fits this model where $\Lambda$ is the string scale energy, $g_*$ is the string coupling, whose small value creates a hierarchy between the string scale, which sets the scale of derivative interactions, and the Planck scale which sets the scale of loop corrections.  \\

The Lagrangian should be Galileon invariant, which means that $L_0$ will take the form \cite{Nicolis:2008in}
\ba
\label{GalileonEFT}
\Lambda^4 L_0 &=& -\frac{1}{2} (\partial \pi)^2 -\frac{1}{2} m^2 \pi^2+ \sum_{n=2}^{4} \alpha_n  \Lambda^3 \, {\cal E} {\cal E} \eta^{4-n} \( \frac{\partial \partial \pi}{\Lambda^3}\)^n \pi \\
&&+ \sum \beta_{p,q}  \Lambda^4  \(\frac{\partial}{\Lambda} \)^p \(\frac{\partial \partial \pi}{\Lambda^3}\)^q \,, \nn
\ea
where the $\alpha_n$ terms are the leading Galileon operators, the $\beta_{p,q}$ are shorthand for all scalar operators built out of contractions of $\partial_{\mu}\partial_{\nu} \pi$ and derivatives thereof,  and we assume $\alpha_n$ and $\beta_{p,q}$ are all order unity or less. As discussed in  \cite{Burrage:2010cu,deRham:2017imi}, in practice the mass term for the Galileon does not break the Galileon symmetry and so this represents a very natural IR extension of the massless Galileon.
 The leading $\alpha_n$ interactions are written in shorthand where, for example,
\be
\alpha_2  \Lambda^3 \, {\cal E} {\cal E} \eta^{2} \( \frac{\partial \partial \pi}{\Lambda^3}\)^2 \pi \quad \text{ means }\quad  \alpha_2  \Lambda^3 \, {\cal E}^{abcd} {\cal E}^{ABCD} \eta_{aA}\eta_{bB} \( \frac{\partial_c \partial_C \pi}{\Lambda^3}\)\( \frac{\partial_d \partial_D \pi}{\Lambda^3}\) \pi \, ,
\ee
i.e. where $\eta$ and $\partial \partial \pi$ are contracted between the pairs of Levi-Civita symbols.

\subsection{Coupling to Matter and Vainshtein}

The Vainshtein mechanism is determined by the coupling to matter, and to be consistent with how the Galileon arises in massive theories of gravity it is the canonically normalized $\pi_c$ that couples to $T$ with $1/\mpl$ strength, and so in the above parameterization the coupling is
\be
\frac{1}{g_* \mpl} \pi T\,.
\ee
Alternatively, canonically normalizing $\pi = g  \pi_c$ then the leading action is
\ba
\label{eq:LEFT}
{\cal L} = -\frac{1}{2} (\partial \pi_c)^2 -\frac{1}{2} m^2 \pi_c^2+ \sum_{n=2}^{4} \alpha_n  \Lambda_*^3 \, {\cal E} {\cal E} \eta^{4-n} \( \frac{\partial \partial \pi_c}{\Lambda_*^3}\)^n \pi_c \\
+ \sum \beta_{p,q}  \frac{\Lambda_*^6 }{\Lambda^2} \(\frac{\partial}{\Lambda} \)^p \(\frac{\partial \partial \pi_c}{\Lambda_*^3}\)^q + \frac{1}{ \mpl } \pi_c T + \dots\nn\,,
\ea
where we have defined
\be
\Lambda_*^3 = \frac{\Lambda^3}{g_*} \, .
\ee
The Vainshtein mechanism kicks in when the value of $\partial \partial \pi \ge \Lambda^3$, or equivalently when $\partial \partial \pi_c \ge \Lambda_*^3$. At this point the leading $\alpha_n$ Galileon operators renormalize the kinetic term for $\pi$, and as long as $\partial  \partial \partial \pi \ll \Lambda^4$ the $\beta_{p,q}$ terms are all parametrically smaller even if $\beta_{p,q} \sim 1$ since they contain at least two extra derivatives' suppression, $\partial^2/\Lambda^2 \sim 1/(\Lambda^2 r_V^2)\ll 1$.  In the full picture of the Vainshtein resummation mechanism, it is assumed we can push deep into the region where $\partial \partial \pi \gg \Lambda^3$ and the $\beta$ terms in the second line of \eqref{eq:LEFT} organize themselves into an expansion such that their magnitude is still suppressed relative to the $\alpha$ terms, just because of the overall derivative suppression. This could occur by a special resummation of the $\partial \partial \pi $ structure, or simply by imagining that all the terms $\beta_{p,q}$ with $q>3$ are suppressed. A concrete example of the former is given in section~\ref{sec:EnteringVain}. We also draw a parallel with the situation in GR with string corrections arising below the Planck scale in section~\ref{sec:GR}. \\

More precisely, the Vainshtein region is defined at distances $r<r_V$, or energies which satisfy $E>1/r_V$, where the Vainshtein radius $r_V$ is determined by the requirement that $\partial \partial \pi_c \sim \Lambda_*^3$, which for a source of mass $M_0$ is when
\be
\frac{M_0}{r_V^3 \mpl } \sim \Lambda_*^3\,,
\ee
or when
\be
r_V = \frac{ M_0^{1/3} }{\mpl ^{1/3}\Lambda_*}.
\ee
To be absolutely sure that we can describe the Vainshtein region within the regime of validity of the EFT, we only require that
$r_V^{-1} \ll \Lambda$, since in the worst case scenario we expect new states of mass $\Lambda$ to arise in the contribution to the forces at these distances.\\

Unlike as assumed for instance in Ref.~\cite{Bellazzini:2017fep}, there is no requirement that $\Lambda \gg \Lambda_*$, i.e. {\bf there is no requirement that $g_* \gg 1$} since the smallness of the $\beta$ terms to the $\alpha$ terms is determined by the hierarchy between $1/r_V$ and $\Lambda$,  {\bf not the hierarchy between $\Lambda$ and $\Lambda_*$}.

\subsection{Improved Positivity Bound and Weak Coupling}

The hierarchy $r_V^{-1} \ll \Lambda$ can easily be achieved while maintaining $g_*\ll 1$, since it requires that
\be
\Lambda_* \left( \frac{\mpl }{M_0} \right)^{1/3} \ll \Lambda = g_*^{1/3} \Lambda_*
\ee
so therefore, 
\be
\left( \frac{\mpl }{M_0} \right) \ll g_* \ll 1 \, .
\ee
Since the mass $M_0$ of most astrophysical objects dominating the Vainshtein screening (e.g. Earth, Sun) is very large in Planck units, this double hierarchy is easy to achieve.\\

Now if we were to factor in the improved positivity bound (equation (\ref{bound1}) for the massive Galileon \cite{deRham:2017imi}) obtained by matching to the IR cross section, we would have a bound on the mass like
\be
\label{Galileonbound}
 \frac{g_*^2m^2}{\Lambda^6} \gtrapprox \frac{g_*^4}{\Lambda^4} \epsilon^8  \,,
 \quad \text{  i.e.        }  \quad \quad m^2 \gtrapprox g_*^2 \Lambda^2 \epsilon^8 = g_*^{8/3} \Lambda_*^2 \epsilon^8 \,,
\ee
where we have neglected order unity factors for simplicity. Here $\epsilon$ enters via the integral over the IR cross section with the maximum taken at $\mu = E^2=\epsilon^2 \Lambda^2$. \\

In the worst case scenario, let us take $\epsilon \sim 1$\footnote{Including factors of $\epsilon$, e.g. $\epsilon \sim 10^{-1}$, does not substantially change any of the following conclusions, it only allows $g_*$ to be slightly larger.} and suppose that we just saturate this bound. Then we may use it to infer $g_*$, 
\be
g_* = \left( \frac{m}{\Lambda_*}\right)^{3/4} \,.
\ee
With the normal `massive gravity' expectation for the scale $\Lambda_*$, namely $\Lambda_*^3 = m^2 \mpl $, then
\be
g_* = \left( \frac{m}{\mpl }\right)^{1/4}  \,.
\ee
The requirement of the existence of the Vainshtein regime described within the validity of the low energy EFT is then
\be
\left( \frac{\mpl }{M_0} \right) \ll  \left( \frac{m}{\mpl }\right)^{1/4}\,.
\ee
The scale $\Lambda$ at which derivative corrections might be expected to kick in is
\be
\Lambda \sim g_*^{1/3} \Lambda_* = \left( \frac{m}{\mpl }\right)^{1/12} \Lambda_*\,.
\ee
Since in the usual assumption $m\sim 10^{-32}$ eV,
\be
\left( m/\mpl  \right) \sim  10^{-60}\,,
\ee
and so $1/\Lambda \sim 10^{5} /\Lambda_*$. This is clearly much worse than normally assumed, but what is important is that there is still a region in which the Vainshtein mechanism takes place and we have not reached the scale $\Lambda$, since we can still satisfy $r_V^{-1} \ll \Lambda$, and therefore even within the third approach, there is still room for a standard Wilsonian UV completion while entering the Vainshtein mechanism.\\

With the standard choice that $m$ is order Hubble scale or so,  then $1/\Lambda_*$ is of the order 1000 km, and so $1/\Lambda \sim 100 {\text{ million km}}$. This is about the distance between the Sun and Venus, and so the EFT could be used to describe the solar system from Venus out, which is all well inside the Vainshtein radius (typically kpc for the Sun). \\

The fact that we have measured with good precision the trajectory of Mercury around the Sun does not rule out this low energy EFT as a valid description of gravity at distances from Venus onwards (i.e. most of the solar system and all the way up to cosmological scales). It just means we need to find a partial UV completion that can push to higher energy. Moreover, the coupling of the Galileon to matter is suppressed by at least 8 orders of magnitude when entering the region where a completion is required within the third approach, and the force mediated by the Galileon is hence 8 orders of magnitude smaller than that of GR:
\be
\frac{F_{\text{Fifth Force}}(r=\Lambda^{-1})}{F_{\rm GR}(r=\Lambda^{-1})} \sim \(\frac{r}{r_V}\)^{3/2}\sim \(\frac{\Lambda_*}{\Lambda}\)^{3/2} \sim 10^{-8} \, ,
\ee
which illustrates the fact that while the Galileon has large self-interactions, it lives in a sector that decouples from the tensor modes and the Standard Model.  Expecting that a UV completion of Galileons would reverse this screening and start suddenly enhancing the force mediated by the Galileon to a point which would be observable by current tests of GR would require a very particular tuning. There is indeed no indication that the relative force mediated by the Galileon would suddenly get enhanced when considering the strongly coupled region. Quite the opposite, since it is highly unlikely that a strongly self-interacting theory can propagate long range forces.
The fact that we have tested GR outside the regime of validity of the standard Galileon EFT by no means  implies that observations rule out the Galileon EFT altogether. As we have emphasized, all it implies is that a better description of the system is required to fully understand its behaviour beyond the EFT regime (if one follows a standard weakly coupled EFT picture with the requirement of a standard Wilsonian UV completion in which the terms which arise beyond the decoupling limit can be used to respect the positivity bounds).

\subsection{Vainshtein resummation}

The usual assumption of the Vainshtein mechanism is that the $\beta$ terms on the second line of \eqref{eq:LEFT} resum around a background configuration $\bar \pi$, for which the leading $\alpha$ terms receive a kinetic renormalization $Z^{\mu\nu}(\bar \pi)$, so that the scale at which derivative interactions enter is now $\Lambda_c=\sqrt{Z} \Lambda$. This is then the cutoff in the presence of a background. This is a highly nontrivial assumption about the UV completion, but by no means an impossible one. \\

There are a couple of ways that we might imagine this to happen. For example, if the masses of all the heavy fields (collectively symbolised here by a single scalar $H$) scale with $Z$, i.e. in the form
\be
{\cal L}_H \sim   \frac{1}{g_*^2} \left( -\frac{1}{2}\partial_{\mu} H \partial_{\nu} H \eta^{\mu\nu} - \frac{1}{2}\Lambda^2 \( 1+ {\rm Tr}[Z]\)  H^2  + \lambda H \Box \pi \dots \right)\,.
\ee
then it is clear that integrating out $H$ already at tree level will generate interactions with the natural derivative suppression $\frac{1}{\Lambda \sqrt{Z}} \partial$ for $Z \gg 1$. Furthermore this property will be preserved by loops.  \\

Alternatively, we can imagine that the massive states which are integrated out to generate the $\beta$ operators have kinetic terms governed by the $Z^{\mu\nu}$ inferred from the leading $\alpha$ terms, e.g. that the heavy particles $H$ couple via a tree level action of the form
\be
{\cal L}_H \sim   \frac{1}{g_*^2} \sqrt{{\rm Det Z}}\left( -\frac{1}{2}\partial_{\mu} H \partial_{\nu} H (Z^{-1})^{\mu\nu}(\pi) - \frac{1}{2}\Lambda^2 H^2  \dots \right)\,.
\ee
In such an example, integrating out $H$ at loop level will clearly generate $\beta$ terms that automatically restructure themselves around a background so that the derivative expansion is controlled by $\partial \partial  /(\bar Z \Lambda^2)  $, where $\bar Z$ denotes the background value of this order parameter.
\\

In the case of a cubic Galileon, in the Vainshtein region we have $Z \sim \partial \partial \pi_c/\Lambda_*^3 \sim (r_V/r)^{3/2}$.
The region in which we can describe the Vainshtein screening within the validity of the low energy EFT is then
\be
\frac{1}{r} \ll \sqrt{Z} \Lambda = (r_V/r)^{3/4} \Lambda\,,
\ee
which we can rewrite as
\be
r \gg   \frac{\Lambda^{-1}}{(\Lambda r_V)^3}\,.
\ee
This is easy to satisfy given a large hierarchy between $r_V$ and $1/\Lambda$.\\

For illustrative purposes, we can estimate the resummed cutoff $\Lambda_c$ at the surface of the Earth, with now the Earth as the main screener. One finds the usual answer of $1$cm, rescaled by $1/g_*^{1/3}$,
\be
\Lambda_c =(r_V/r)^{3/4} \Lambda=(r_V/r)^{3/4} \Lambda_* g_*^{1/3} = \(1{\rm cm}/g_*^{1/3}\)^{-1}
\ee
i.e. a distance scale of
\be
1 {\rm cm} \times \left( \frac{\mpl }{m}\right)^{1/12} \sim 800 \, {\rm m}\,,
\ee
where we have made the usual assumption that the mass is of order the Hubble scale. This means that the EFT can be used to describe the Earth--Moon orbit, but will break down at distances of order km, so we cannot in principle use it to describe table-top experiments\footnote{In itself, this is not technically correct, since for any experiment the local environment needs to be included and in turn redresses the scale at which the EFT breaks down \cite{Berezhiani:2013dca}.}. However, this certainly does not rule out massive Galileons (and by extension massive gravity) as an IR modified theory valid at solar system and cosmological scales---it simply means that we need to find a (partial) UV completion to describe local physics. To actually rule out an EFT, one would have to show that it gave a wrong answer for a physical observable in the regime we can trust the EFT, e.g. for lunar laser ranging and the Earth--Moon orbit.\\

Moreover, as emphasized earlier, the virtue of the Vainshtein mechanism is to entirely decouple a sector (in this case the Galileon to the Standard Model), so while at distances shorter than the order of km the Galileon EFT breaks down, it does so in a sector which becomes increasingly decoupled from observables. The couplings of the Galileon to matter is indeed suppressed by 5 orders of magnitude as compared to GR.  Once again,  it is highly unlikely that the strong coupling effects that would occur in this theory at shorter distances would suddenly revert this suppression, as strongly coupled systems typically do not propagate long-range forces.

\subsection{Vainshtein analogue in GR/string theory}
\label{sec:GR}

Although the Vainshtein region, $\partial \partial \pi \gtrapprox \Lambda^3$, is na\"{i}vely outside of the regime of validity of the EFT expansion, it is helpful to compare this with a well understood example where a similar situation arises: namely GR with new physics at a scale below the Planck scale. A good example is the low energy effective action for weakly coupled string theory, which takes the form (focusing on only the spin 2 terms and fixing the value of the dilaton)
\be
\label{STEFT2}
S = \frac{1}{g_s^2}  \int \d^ {10} x \sqrt{-g}\(  \Lambda^8 R + \Lambda^{10} \sum \gamma_{p,q} \( \frac{\nabla}{\Lambda}\)^p \( \frac{R_{\mu\nu\rho\sigma}}{\Lambda^2} \)^q + \dots  \) \,,
\ee
where $1/\Lambda=\sqrt{\alpha'}$ is the string length and the 10 dimensional Planck scale is identified as
\be
\mpl^8 = \Lambda^8/g_s^2 \, .
\ee
The $\gamma_{p,q}$ terms in \eqref{STEFT2}  are the $\alpha'$ or string scale corrections that arise from integrating out the excited states of the string which are coupled to the massless graviton mode.
The precise structure of these  terms is determined by the low energy expansion of the tree level stringy graviton scattering amplitudes. \\

Further expanding perturbatively, $g_{\mu\nu}= \eta_{\mu\nu}+ \frac{h_{\mu\nu}}{\mpl^4}$, then the effective action takes the form
\ba
\label{STEFT}
S =  \int \d^ {10} x \, \Bigg(  -\frac{1}{2} h \hat{\mathcal{E}} h  + \mpl^{8} \Lambda^2 \sum_{p} \alpha_p \( \frac{h}{\mpl^4}\)^p \(\frac{\partial h}{\Lambda \mpl^4}\) \(\frac{\partial h}{\Lambda \mpl^4}\) \\
+ \mpl^8 \Lambda^{2} \sum_{p,q}   \beta_{p,q} \( \frac{\partial}{\Lambda}\)^p \left(\frac{h}{\mpl^4} \right)^q \Bigg) +  \dots\nn\,,
\ea
where $\hat{\mathcal{E}}$ is the Lichnerowitz operator in 10 dimensions, the $\alpha_p$ terms come from expanding the Einstein-Hilbert term, and the $\beta_{p,q}$ terms from expanding the $\alpha'$ corrections (the $\gamma_{p,q}$ corrections of \eqref{STEFT2}).  This closely parallels the form of the massive Galileon EFT \eqref{GalileonEFT}, or the massive gravity EFT considered later \eqref{MGEFT}. In this analogy, $h/\mpl^4$ plays the role of $\partial \partial \pi^c/\Lambda_3^3$ and the Schwarzschild radius plays the role of the Vainshtein radius. \\

From a na\"{i}ve EFT point of view, looking at (\ref{STEFT}), the expansion will break down when $h \sim \mpl^4$, which occurs for instance in the vicinity of a black hole horizon, and the $\beta $ terms for a sufficiently large power of $h$ could swamp the $\alpha$ terms, rendering the EFT out of control.  However, this is not the case, as this theory exhibits an analogous `Vainshtein mechanism' as follows: in the vicinity of a source for which $h \gg \mpl^4$, provided the sources varies over distance scales larger than $1/\Lambda$ (i.e. $\partial \ll \Lambda$), the background geometry remains under control with the $\beta$ terms remaining negligible. This is made transparent in the form (\ref{STEFT2}), since there it is clear that as long as the curvature is small compared to the string scale, $R \ll \Lambda^2$, and as long as derivatives are small, $\nabla \ll \Lambda$, solutions with $h \sim \mpl^4$ or even $h \gg \mpl^4$ will remain under control in the EFT.  The key point here is that the special resummation of the EFT expansion required for this to work is ensured by diffeomorphism invariance, and because of this symmetry the EFT automatically reorganizes itself as a new EFT expansion around a background solution even for $h \gg \mpl^4$.
Furthermore, {\bf this `Vainshtein mechanism' is perfectly consistent with weak coupling $g_* \ll 1$}, as we certainly have no problem describing black holes in weakly coupled string theory! There is even an analogue of the Vainshtein resummation of the cutoff here, because on a curved background the breakdown of the derivative expansion is not set by $\partial_{\mu} \sim \Lambda$, but rather by the locally inertial value of $\nabla \sim \Lambda$, i.e. $e^{\mu}_a \nabla_{\mu} \sim \Lambda$ where $e^{\mu}_a $ is the inverse vielbein. Here the inverse vielbein $e^{-1}\sim 1/\sqrt{g}$ plays the same role as $1/\sqrt{Z}$. Thus the Vainshtein resummation is guaranteed by virtue of the equivalence principle.
The successful description of the Vainshtein region in theories of massive Galileons and massive gravity requires some analogous resummation structure for $\partial \partial \pi/\Lambda^3$ in the former and $K_{\mu \nu}$ (see section \ref{sec:EnteringVain}) in the latter.

\section{Weakly Coupled Massive Gravity}
\label{massivegravity}

\subsection{Connection with Galileons}

The extension of this discussion to massive gravity is relatively straightforward once one recognizes that, in the limit $m \ll \mpl $, the dominant interactions in massive gravity are determined by the helicity-zero mode, which is effectively a massive Galileon.
Working in the \stu formulation, with 4 \stu fields $\phi^a$ to account for the broken diffeomorphisms, then defining the tensor
\ba
K\mupn = \delta\mupn - \sqrt{g^{\mu \alpha} \partial_{\alpha} \phi^a \partial_{\nu} \phi^b \eta_{ab}}\, ,
\ea
we can denote the leading order massive gravity action as
\be
S_{\rm leading} = \int \d^4 x \sqrt{-g} \frac{1}{g_*^2} \left[  \frac{M^2}{2} R  - m^2 M^2 \sum_{n} \alpha_n {\cal E} {\cal E} g^{4-n} K^n \right]\,,
\ee
where the mass scale $M$ is such that
\be
M^2  = g_*^2 \mpl ^2\, ,
\ee
while $\alpha_0=\alpha_1=0$ to remove tadpole terms/cosmological constant, and $\alpha_2$ is chosen so that the term quadratic in $K$ has the correct Fierz-Pauli mass normalization,
\be
\left[ \frac{M^2}{2} R  - \frac{m^2 M^2}{2} (K\mupn K^\nu_{\, \mu} - (K^{\mu}_{\, \mu})^2) + \dots \right] \, .
\ee
For these leading interactions, there are then two free parameters: $\alpha_3$ and $\alpha_4$.
Weak coupling in the helicity-2 sector is the statement that there is a new energy scale, $M$, which is parametrically lower than $\mpl $, at which we expect curvature corrections to kick in.
To maintain the $1/g_*^2$ normalization of the action, the \stu fields are naturally expanded as
\be
\phi^a = x^a - \frac{V^a}{m M}  - \frac{\partial^a \pi}{m^2 M} \, , \quad g_{\mu\nu}= \eta_{\mu\nu} + \frac{h_{\mu\nu}}{M} \, .
\ee
Alternatively, in canonical normalization we have
\be
\phi^a = x^a - \frac{V_c^a}{m \mpl }  - \frac{\partial^a \pi_c}{m^2 \mpl } \, , \quad g_{\mu\nu}= \eta_{\mu\nu} + \frac{h^c_{\mu\nu}}{\mpl } \, ,
\ee
reflecting the fact that for all fields $\Phi$, the canonical normalization is $\Phi = g_* \Phi_c$.
In the limit $m \ll \mpl $,
\ba
K\mupn  \rightarrow \frac{\partial^{\mu} \partial_{\nu} \pi}{m^2 M} + \cdots=\frac{\partial^{\mu} \partial_{\nu} \pi_c}{m^2 \mpl} + \cdots \, ,
\ea
where the ellipses indicate corrections that vanish in the decoupling limit, i.e. corrections that are suppressed by a scale higher than $\Lambda^3 = m^2 M$. The new scale $\Lambda$ connects the scale of new physics in the helicity-2 sector with that in the helicity-0 sector. \\

\subsubsection{Identification of the helicity-0 and -1 modes in all generality}

In conventional treatments, it is stated that only in the high energy limit, $E \gg m$, can $\pi$ and $V^a$ be identified as the helicity-zero and helicity-one modes of the massive spin-2 field (Goldstone equivalence theorem) and so this decomposition is only useful there. We can however do better than this, by adding Fadeev-Popov gauge fixing terms which diagonalize the `St\"uckelberg-ized' quadratic (Fierz-Pauli) lagrangian even at low energies. Specifically, to the Fierz-Pauli Lagrangian,
\ba
{\cal L}_{\rm FP} &=& \frac{1}{8 g_*^2} \left[ h^{\mu\nu} \Box(h_{\mu\nu}- \frac{1}{2} h \eta_{\mu\nu}) + 2 (\partial^{\mu} (h_{\mu\nu}-\frac{1}{2} h \eta_{\mu\nu}))^2 \right. \\
&& \left. -m^2\(  \({h_{\mu\nu} + \frac{1}{m} \partial_{\mu} V_{\nu}+ \frac{1}{m} \partial_{\nu} V_{\mu} + \frac{2}{m^2} \partial_{\mu} \partial_{\nu} \pi}\) ^2-\(h+ \frac{2}{m} \partial V + 2 \frac{\Box}{m^2}\pi \)^2 \) \right]\,, \nn
\ea
we may add the following diffeomorphism and $U(1)$ gauge fixing terms,
\be
{\cal L}_{\rm GF} = -\frac{1}{8 g_*^2}\(  2 ( \partial^{\mu} (h_{\mu\nu}-\frac{1}{2} h \eta_{\mu\nu}) + m V_{\mu} +  \partial_{\mu} \pi )^2  + \frac{1}{2} ( 2 \partial_{\mu} V^{\mu} + m h + 2 m \pi)^2 \)\,.
\ee
Then defining $\tilde h_{\mu\nu} = h_{\mu\nu}- \pi \eta_{\mu \nu}$, the quadratic Lagrangian is fully diagonal,
\be
{\cal L}_{\rm FP} +{\cal L}_{\rm GF}  = \frac{1}{8 g_*^2} \left[ \tilde h^{\mu\nu} (\Box-m^2)(\tilde h_{\mu\nu}- \frac{1}{2} \tilde h \eta_{\mu\nu})  + 2 V^{\mu} (\Box-m^2) V_{\mu} + 6\pi (\Box-m^2) \pi \right]\,,
\ee
meaning that the polarization structure for all propagators is trivial (that of a scalar). By adding an appropriate nonlinear gauge fixing term which has this form at the quadratic level, we make manifest the fact that the scattering amplitudes will contain the interactions of a massive Galileon, together with the additional `vector' $V_{\mu}$ and `tensor' $\tilde h_{\mu\nu}$ degrees of freedom. This diagonalization of the off-shell degrees of freedom, and the removal of all second class constraints and nontrivial polarization structure from the propagators makes it significantly more straightforward to count the size of the EFT corrections, even at low energies. This can be used to give an off-shell meaning to the SVT decomposition, of which the on-shell amplitudes are projections. \\

In terms of the diagonalized degrees of freedom, the linearized gauges  are
\be
 F_{\mu}  = \partial^{\mu} (\tilde h_{\mu\nu}-\frac{1}{2} \tilde h \eta_{\mu\nu}) + m V_{\mu} =0 \, ,  \quad  F_{\pi}  = 2 \partial_{\mu} V^{\mu} + m \tilde h -  2 m \pi =0 \, .
\ee
By construction, in the limit $m\rightarrow 0$, these gauge choices degenerate into the standard de Donder (harmonic) and Lorenz gauges for massless spin 2 and spin 1 fields, as expected from the identification of the modes at high energy with the massless helicity states. On-shell polarization vectors must be chosen to satisfy these gauge conditions, and we see that when $m \neq 0$ there is a mixing between the off-shell SVT modes. \\

Under a linear diffeomorphism: $h_{\mu\nu} \rightarrow h_{\mu\nu}+ \partial_{\mu} \xi_{\nu}+ \partial_{\nu} \xi_{\mu}$, $V_{\mu} \rightarrow V_{\mu}- m \xi_{\mu}$, $\pi \rightarrow \pi$, and then  $F_{\mu}$ and $F_{\pi}$ transform as
\be
F_{\mu} \rightarrow F_{\mu} + (\Box-m^2) \xi_{\mu} \, , \quad F_{\pi} \rightarrow F_{\pi} .
\ee
Under a $U(1)$ transformation: $h_{\mu\nu} \rightarrow h_{\mu\nu}$, $A_{\mu} \rightarrow A_{\mu} + \partial_{\mu} \xi $, $\pi \rightarrow \pi - m \xi$, so then,
\be
F_{\mu} \rightarrow F_{\mu} \, \quad F_{\pi} \rightarrow F_{\pi} + 2 (\Box -m^2) \xi \, .
\ee
It is thus apparent that in this gauge, the Fadeev-Popov ghosts\footnote{Of course, quantizing a theory diffeomorphism invariance is more subtle than the usual Fadeev-Popov procedure, but we will not dwell on these issues here. The $U(1)$ part is at least conventional.} are also diagonalized particles of the same mass $m$. Indeed the gauge fixing terms can be determined on this basis alone. As in standard quantization in Lorentz invariant gauges, we still have the freedom to perform gauge transformations $(\xi^{\mu},\xi)$ which satisfy
\be
[\Box-m^2] \xi^{\mu} = [\Box-m^2] \xi =0 \, ,
\ee
which can be used to remove the unphysical longitudinal modes in the asymptotic states.
Specifically, working in a rest frame for a graviton of momentum $k^{\mu}=(m,0,0,0)$ we can solve $F_{\pi}=0$ to determine $V_0$ and $F_{\mu}=0$ to determine $h_{0 \mu}-\frac{1}{2} \eta_{0\mu} h$. We can then use the remaining gauge freedom to set $\partial_i V^i = \partial^k h_{kj} =h^i_i =0 $. The resulting transverse and traceless $h_{ij}$, transverse $V_i$, and $\pi$ are then the physical on-shell SVT degrees of freedom.
\\

A particular choice of nonlinear gauges  which is consistent with the linear gauge choice is
\ba
&& F^a = M \( - (\Box_g -m^2) x^a -m^2 \phi^a \) \, , \\
&& F_{\pi} =-2 m M \partial_a \phi^a - \frac{2}{m} ( \Box_{\eta}-m^2) \pi+ 4 m M + m M \eta^{\mu\nu} g_{\mu\nu} \, ,
\ea
where $\Box_g$ is the covariant d'Alembertian and $\Box_{\eta}$ the Minkowski d'Alembertian. This pair have been chosen so that the tree level vertices and propagators determined from $\Lambda^4 L_0 + {\cal L}_{\rm GF}$ with
\be
 \sqrt{-g}{\cal L}_{\rm GF} = -\frac{1}{8 g_*^2} \left( \sqrt{-g} \(2 g_{ab}F^a F^b \)+ \frac{1}{2} F_\pi^2 \)\,,
\ee
continue to respect the Galileon symmetry (realized in the sense that $\pi \rightarrow \pi + v_{\mu} x^{\mu}+c$ and $V^a \rightarrow V^a - v^a/m$ when perturbed around an arbitrary background). Indeed, as in the case of the massive Galileon \cite{deRham:2017imi}, the terms that violate the symmetry are purely quadratic, and hence all tree level vertices and propagators naturally respect the symmetry. Alternative gauge choices typically break the Galileon symmetry, but only through $m/\mpl$ suppressed terms.

\subsubsection{Wilsonian Effective action for Weakly Coupled Massive Gravity}

Within the framework of approach \ref{appro3}, the general structure of a `single scale  - single coupling' tree level Lagrangian for weakly coupled massive gravity is \cite{deRham:2010kj}
 \be
{\cal L} = \frac{1}{g_*^2}\left( \Lambda^4 L_0 + g_*^2 L_1 + \dots \right)\,,
\ee
where
\be
\label{MGEFT}
\Lambda^4 L_0 =  \left[  \frac{M^2}{2} R  - \Lambda^3 M \sum_{n} \alpha_n {\cal E} {\cal E} g^{4-n} K^n \right] + \Lambda^4 \sum \beta_{p,q,r} \( \frac{\nabla}{\Lambda}\)^p K_{\mu\nu}^{q} \( \frac{R_{\mu\nu \rho \sigma}}{\Lambda^2} \)^r \, .
\ee
The form of the interactions in $L_1$ and higher will be those needed to renormalize loops from $L_0$.
We have included the possibility that the curvature corrections come in at the scale $\Lambda$ since such terms will arise from the commutator of two covariant derivatives, however we expect that those curvature corrections that relate to the helicity 2 sector, i.e. those present in the absence of the mass term, will be suppressed by some higher scale, e.g.  $M^2$, 
\be
\Lambda^4 \Delta L_0 \sim M^4 \( \frac{\nabla}{M}\)^a \( \frac{R_{\mu\nu \rho \sigma}}{M^2} \)^b \, .
\ee
Having in mind the hierarchy $\Lambda \ll M$, such interactions will be irrelevant to the following considerations.
An important point to recognize about an approach \ref{appro3} UV completion is that the terms $\beta_{p,q,r}$ will necessarily be non-zero in order to respect the higher order positivity bounds (\ref{generalbounds}) (this was discussed in the massive Galileon case in \cite{deRham:2017imi}). This is not necessarily the case for an approach \ref{appro1} UV completion, where the ghost-free structure of the leading interactions is a more important feature. \\

The EFT expectation is that the coefficients $\alpha_n$ and $\beta_{p,q,r}$ are all at most of order unity, although individual ones may be much smaller as a consequence of special properties of the UV theory.
In terms of the perturbative fields, the corrections to the leading order action take the form
\be
\Lambda^4 \sum \beta_{p,q,r} \( \frac{\nabla}{\Lambda}\)^p K_{\mu\nu}^{q} \( \frac{R_{\mu\nu \rho \sigma}}{\Lambda^2} \)^r  \sim \Lambda^4 \sum \gamma_{lmnp} \(\frac{\partial}{\Lambda}\)^l  \(\frac{\partial \partial \pi}{\Lambda^3}\)^m \(\frac{m \partial V}{\Lambda^3}\)^n \(  \frac{\tilde h}{M} +\frac{m^2 \pi}{\Lambda^3}\)^p \, .\nn
\ee
The fact that the interactions of $V$ and $h$ come in at a different scale than $\pi$ is fixed by a combination of diffeomorphism invariance and the $U(1)$ gauge symmetry, so is preserved under loops.
For instance, under the $U(1)$ gauge symmetry, it is $D_{\mu} \pi = \partial_{\mu} \pi + m V_{\mu}$ which is gauge invariant. \\

Due to the total derivative nature of the ${\cal E} {\cal E} g^{4-n} K^{n}$ at leading order in the decoupling limit, the $\alpha$ terms have the structure
\be
- \Lambda^3 M \sum_{n} \alpha_n {\cal E} {\cal E} g^{4-n} K^n \sim  \Lambda^3 M  \(\frac{\partial \partial \pi}{\Lambda^3}\)^{m} \(\frac{m \partial V}{\Lambda^3}\)^n \( \frac{\tilde h}{M} +\frac{m^2 \pi}{\Lambda^3}\)^{p} \, ,
\ee
and the tuned `ghost-free' structure ensures that $n \ge 2$ when $p=0$.
The overall counting of these terms is different due to their different structure, but this structure is preserved due to the combination of diffeomorphisms and the $U(1)$ symmetry.
Note that for both sets of interactions, the terms that arise from the diagonalization $h \rightarrow \tilde h$ are of order $m^2 \pi /\Lambda^3$ and terms of the same order come from the IR part of $\partial \partial \pi/\Lambda^3$ when $\partial \sim m$. Thus, if we are only interested in counting the order of magnitude of terms in the EFT expansion, we may ignore these particular interactions, unless the others happen to vanish. This will occur for certain terms from the leading interactions. \\

According to this off-shell split, the interactions from the Einstein-Hilbert term only affect the off-shell tensors, for instance at tree level for the $2 \rightarrow 2$ amplitude it will only occur through the $h h (\partial h ) \partial h/M^2$ interaction in the $TTTT$ amplitude and the $ h (\partial h ) (\partial h)/M$ at one vertex in the $TXTX$ amplitude.
On-shell there is some mixing into the other amplitudes due to the $m/\sqrt{s}$ suppressed projections. Regardless, these interactions will be suppressed relative to the leading ones.
Similarly those interactions arising from ${\cal L}_{\rm GF}$ are all at least $\mpl$ suppressed, and in certain cases additionally $m/\mpl$ suppressed, and so may similarly be neglected relative to the leading contributions. This is a consequence of choosing a gauge with a good behaviour in the $m\rightarrow 0 $ limit. \\

\subsection{EFT counting for the Scattering Amplitude: Leading Interactions}

Let us now consider a $2 \rightarrow 2$ scattering amplitude with $n_S$ scalars, $n_V$ vectors and $n_T$ tensors with $n_S+n_V+n_T=4$. The quartic interactions that contribute to the scattering amplitude from the leading terms will be\footnote{As previously discussed, the contributions from the Einstein-Hilbert terms and gauge fixing terms are negligible in the St\"uckelbergized  formalism. This is in stark contrast to the unitary gauge calculation, reflective of the fact that the latter is a poor indicator of the physics. }
\be
 \Lambda^3 M \(\frac{\partial \partial \pi}{\Lambda^3}\)^{n_S-p} \(\frac{m \partial V}{\Lambda^3}\)^{n_V} \( \frac{\tilde h}{M}  \)^{n_T}\( \frac{m^2 \pi}{\Lambda^3} \)^p\,,
\ee
and will contribute to the scattering amplitude in the form
\ba
A(s) &\sim& g_*^2 \Lambda^3 M s^{-2} \(\frac{s^{3/2} }{\Lambda^3}\)^{n_S-p} \(\frac{m s }{\Lambda^3}\)^{n_V} \( \frac{\sqrt{s}}{M} \)^{n_T} \( \frac{m^2 \sqrt{s} }{\Lambda^3} \)^p \nn \\
&\sim & g_*^2 \Lambda^3 M m^{2p+n_V}\frac{s^{3 n_S/2+n_V+n_T/2-2-p}}{\Lambda^{3n_s+3n_V} M^{n_T}}
\sim  g_*^2 \frac{ m^{n_V+2 n_T+2p} s^{4 - n_V/2-n_T-p}}{m^2 \Lambda^{6}} \, ,
\ea
where any individual power of $s$ here is really shorthand for $s, t$ or $m^2$. For instance $s^3$ may denote any combination of $s^3, s^2 t, s t^2, t^3, s^2 m^2 , s t m^2 , t^2 m^2, s m^4 , t m^4,m^6$. Note that the $1/m^2$ which would seem to contradict the existence of a decoupling limit ($m \rightarrow 0 $ with $\Lambda$ fixed) is always cancelled, since if $n_T=p=0$ then $n_V \ge 2$ for these interactions, and so $n_V+2n_T+2p \ge 2$.  \\

It is then straightforward to estimate the contribution of these interactions to the positivity bound,
\be
B^{(2,0)}(s=2m^2)  \sim g_*^2 \frac{m^2}{\Lambda^6} \, .
\ee
Remarkably, all polarizations contribute to the positivity bound at the same order of magnitude, and this term is suppressed by $m^2/\Lambda^2$ relative to the na\"{i}ve expectation. This is, of course, a consequence of the Galileon symmetry giving the scalar and vector amplitude special soft properties, making them comparable to tensor amplitudes.
This is borne out by the explicit unitary gauge calculations first given in \cite{Cheung:2016yqr}. \\

A similar argument can be made for the exchange ($s$, $t$ and $u$ channel) interactions, where we assume that $m_S$ scalars, $m_V$ vectors and $m_T$ tensors are exchanged with $m_S+m_V+m_T=1$. Splitting the number of scalars, vectors and tensors ($P=S,V,T$) at each vertex as $n_P = n_P^L + n^R_P$ then the generic exchange amplitude is\footnote{For $s \gg m^2$ this estimate is strictly true for the $s$ and $u$ channel exchange, whereas for the $t$-channel amplitude evaluated in the forward limit can be enhanced by $s/(m^2-t) \sim s/m^2$. This will not however affect our estimates for the positivity bounds which are evaluated at $s\sim m^2$.}
\ba
A(s) &\sim& g_*^2 \Lambda^6 M^2 s^{-3}  \(\frac{s^{3/2} }{\Lambda^3}\)^{n^L_S+m_S} \(\frac{ms }{\Lambda^3}\)^{n^L_V+m_V} \( \frac{\sqrt{s}}{M} \)^{n^L_T+m_T}  \nn\\
&\times& \frac{1}{s} \(\frac{s^{3/2} }{\Lambda^3}\)^{n^R_S+m_S} \( \frac{m s }{\Lambda^3}\)^{n^R_V+m_V} \( \frac{\sqrt{s}}{M} \nn\)^{n^R_T+m_T}  \\
&\sim& g_*^2 \Lambda^6 M^2 s^{-4}  \(\frac{s^{3/2} }{\Lambda^3}\)^{n_S+2 m_S} \(\frac{m s }{\Lambda^3}\)^{n_V+2 m_V} \( \frac{\sqrt{s}}{M} \)^{n_T+2 m_T} \, .
\ea
The topological constraints that $\sum_P n_P^L+m_P =3$ imply $\sum_P n'_P =6$, where $n_P'=n_P+2m_P$. Imposing the constraint, we have
\be
A(s) \sim g_*^2  \frac{ m^{n'_V+2 n'_T} s^{5 - n'_V/2-n'_T}}{m^4 \Lambda^{6}} \, .
\ee
As before, the $1/m^4$ is always cancelled since $n'_V+2 n'_T \ge n^L_V+ 2 n^L_T+n^R_V+2 n^R_T \ge 4$ and so once again
\be
B^{(2,0)}(s=2m^2)  \sim g_*^2 \frac{m^2}{\Lambda^6} \, ,
\ee
consistent with the known results. \\

The next non-trivial positivity bound comes from $B^{(4,0)}(s=2m^2)$, and this can be similarly estimated as
\be
B^{(4,0)}(s=2m^2)  \sim 0 \, ,
\ee
essentially because the tree level interactions grow at most as $s^3$. This means that we need to go to higher order in either loops or in EFT corrections. \\

A crucial observation was made in \cite{Cheung:2016yqr} that although we receive contributions to $B^{(2,0)}(s=2m^2)$ of the same order, there is a constraint for elastic scattering of any two single particle states which are themselves arbitrary superpositions of polarizations. This means that there are far more constraints from the positivity bounds than there are free coefficients (in the above notation $\alpha_3$ and $\alpha_4$) in the massive gravity Lagrangian. The confluence of these constraints is enough to significantly constraint the parameter space for massive gravity models \cite{Cheung:2016yqr}. It should be stressed however, that this only applies to approach \ref{appro3} UV completions, and there may be no constraints in the parameter space of massive gravity for approach \ref{appro1} and \ref{appro2} UV completions.

\subsection{Improved Positivity Bounds for Weakly Coupled Massive Gravity}

The application of the improved positivity bounds, eqn.~(\ref{bound2}), to the scattering amplitudes of massive gravity has recently been considered in \cite{Bellazzini:2017fep}. Fundamentally, there is no essential difference between the massive gravity case and that of the massive Galileon. For the elastic scattering of a given set of polarizations, $P_1,P_2$, there is a positivity bound which is given in \cite{Cheung:2016yqr} of the form
\be
B^{(2,0)}_{P_1,P_2}(s=2m^2)  = \frac{g_*^2 m^2 }{\Lambda^6} A_{P_1P_2}\,,
\ee
where the coefficients $A_{P_1P_2}$ are dimensionless functions of the two free dimensionless parameters in the leading order interactions: $\alpha_3$ and $\alpha_4$ (or, in the notation of \cite{deRham:2010ik,Cheung:2016yqr,Bellazzini:2017fep}, $c_3$ and $d_5$). The tree level positivity bounds of \cite{Cheung:2016yqr} are that $A_{P_1P_2}>0$.
The improved bounds \eqref{bound2} take the form
\be
\frac{g_*^2 m^2 }{\Lambda^6} A_{P_1P_2} > \frac{g_*^4}{\Lambda^4} \epsilon^8 C_{P_1P_2}\,,
\ee
where $C_{P_1P_2}$ are also order unity functions of $\alpha_3$ and $\alpha_4$ ($c_3$,$d_5$). This is identical in form to the bound for the massive Galileon \eqref{Galileonbound} with the only difference being the multiplicity of bounds from the different polarization states.
Just as in the massive Galileon case, the left-hand side of this inequality vanishes as $m \rightarrow 0$, due to the soft scattering properties associated with the Galileon symmetry, whereas the right hand side is finite as $m \rightarrow 0$ but is additionally suppressed by $g_*^2$. We thus conclude that an approach \ref{appro3} UV completion of massive gravity will similarly require weak coupling $g_* \ll 1$. Given the factor of $\epsilon^8$ on the right hand, then given even the small range of parameters given in \cite{Cheung:2016yqr}, we can safely satisfy the improved inequality if
\be
g_*^2 \le \frac{m^2}{\Lambda^2}\,.
\ee
This is equivalent to
\be
g_* \le \( \frac{m}{\Lambda_*} \)^{3/4}=\( \frac{m}{\mpl} \)^{1/4} \, ,
\ee
and again, choosing the mass scale to be of order the Hubble rate leads to $g_* < 10^{-14}$. Once $g_*$ is in this range, the right hand side is sufficiently negligible that we return to the allowed range of parameters given in \cite{Cheung:2016yqr}, determined by $A_{P_1P_2}>0$. \\

\begin{figure}
\centering
\includegraphics[width=0.8\textwidth]{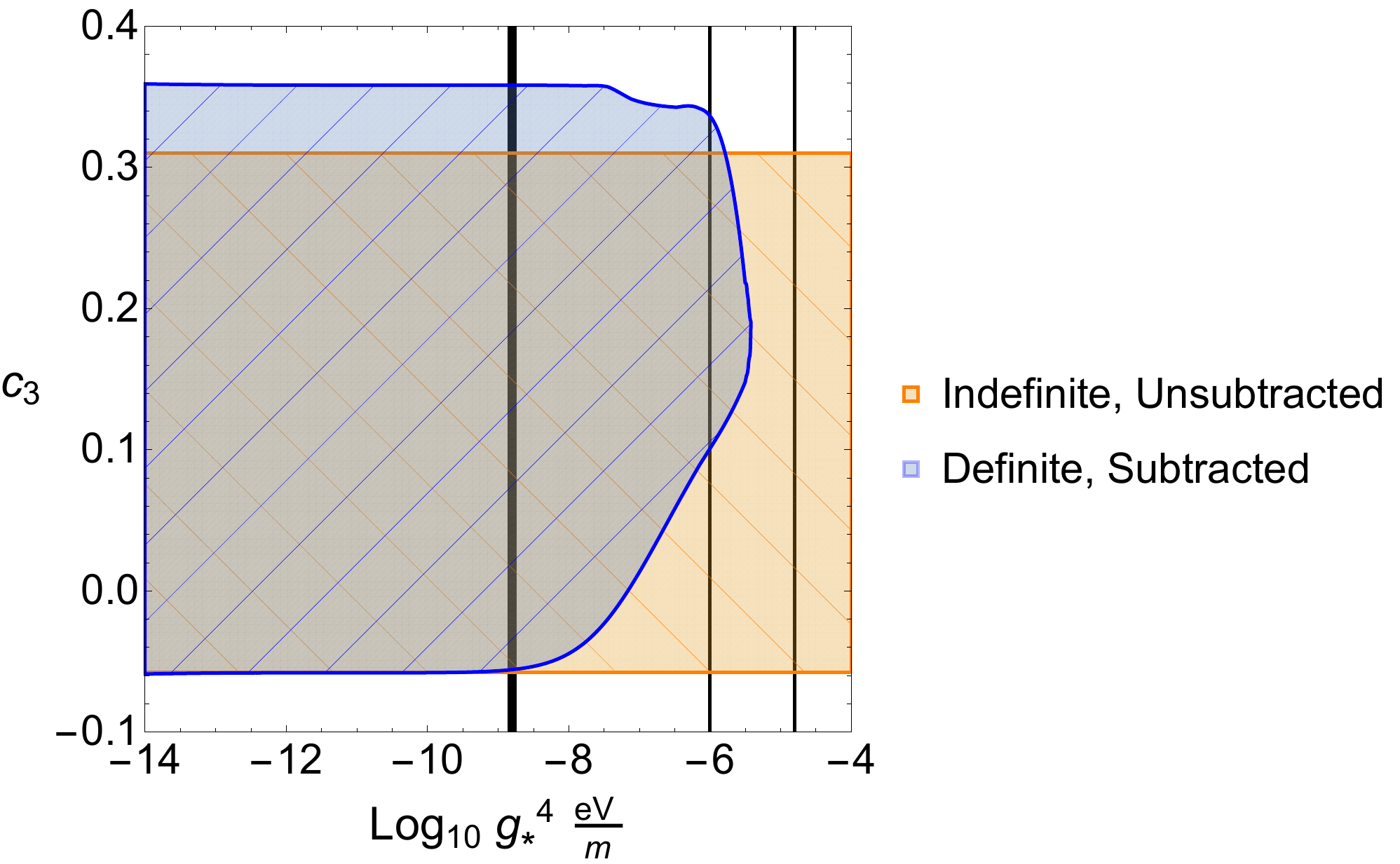}\\[0.5cm]
\includegraphics[width=0.8\textwidth]{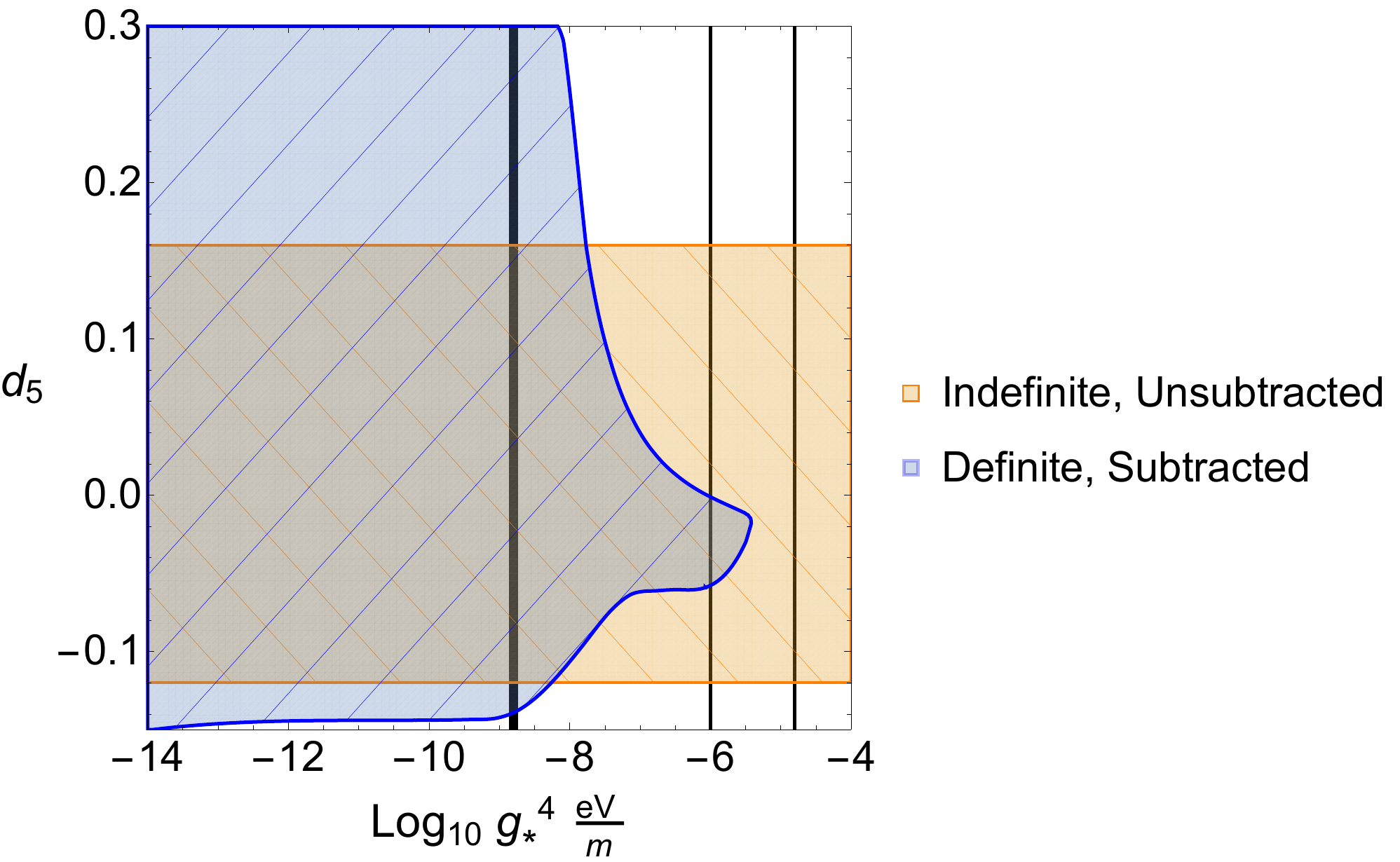}
\caption{Regions allowed in the $\{c_3,d_5\}$ parameter space of massive gravity. Those determined by subtracted bounds from definite SVT scattering (as in \cite{Bellazzini:2017fep}) are provided in
blue, and the unsubtracted bounds from indefinite scattering (as in \cite{Cheung:2016yqr}) are provided in orange. The figure over $c_3$ has been maximised over the allowed values of $d_5$ and vice versa.
 The vertical lines correspond to couplings of $g_*=2 \times 10^{-10}$ (in bold),
and $2 \times 10^{-9}$, $10^{-9}$ (thin lines) with a mass of $m=10^{-30}$eV.
 Ref.~\cite{Bellazzini:2017fep} focused on the two last values of $g_*$ (thin lines), which have no particular physical meaning. As is transparent from the figure, for small
couplings, $g_*^4\lesssim 10^{-8.8} (m/{\rm eV})$,
subtracting part of the cut from SVT amplitudes does not improve
on existing constraints.}
\label{fig:gstar}
\end{figure}

Ref.~\cite{Bellazzini:2017fep} claims that the improved bounds are a vast improvement over that of \cite{Cheung:2016yqr}. However, in reality this is only the case if $g_*$ is tuned to be arbitrarily close to the largest possible value allowed by the inequality. Generically, in the allowed range $g_* \le m^2/\Lambda^2$, there is no significant improvement in the allowed parameter space. We illustrate this statement in Fig~\ref{fig:gstar}. As is clear from the figure, for sufficiently small couplings subtracting part of the cut from SVT amplitudes does not improve on existing constraints. The conclusions of Ref.~\cite{Bellazzini:2017fep} simply relied on tuning the value of $g_*$ to be on the edge of the acceptable region.  \\

Ref.~\cite{Bellazzini:2017fep}  then further makes the specious argument that these improved positivity constraints rule out massive gravity. This is based on the incorrect assumption that the Vainshtein mechanism requires $g_* \gg 1$, which is obviously in conflict with the above requirement. \\

To emphasize once more, the assumption that the Vainshtein mechanism would require $g_* \gg 1$ as stated in Ref.~\cite{Bellazzini:2017fep} is entirely flawed since the Vainshtein mechanism is essentially a classical explanation of the screening of fifth forces and certainly does not require $g_*$ strong coupling\footnote{As already mentioned, when referring in the literature to the need of strong coupling for the Vainshtein mechanism to work, what is meant is  the breaking of perturbative unitarity for irrelevant operators. There has never been any notion that the Vainshtein would ever require $g_*\gg 1$.}, as we have emphasized (its validity only requires $r_V \gg \Lambda^{-1}$). As in the massive Galileon case, it is true that a small $g_*$ will lower the scale at which new physics comes in, since $\Lambda \sim g_*^{1/3} \Lambda_*$, however there is no sense in which this conflicts with observations, it merely limits the applicable regime of the above EFT. Since the Vainshtein mechanism is essentially a decoupling mechanism, by which whatever physics is responsible for the spontaneous breaking of diffeomorphism invariance decouples from the helicity-2 gravity sector and the Standard Model, there is no reason to believe that the UV completion of this dark sector will undo this decoupling even if it becomes strongly interacting at a low scale.

\subsection{EFT corrections from higher derivative operators}

Let us now compute the corrections to the scattering amplitude that arise from the $\beta$ terms in the tree level action to establish the stability of the bounds from the leading order interactions. This is a worthwhile exercise since (a) whilst these are naively suppressed, the leading contribution to $B^{(2,0)}$ is itself suppressed by $m^2/\Lambda^2$ due to the Galileon symmetry and (b) the higher derivative operators are crucial in ensuring higher order positivity bounds are satisfied. \\

From quartic interactions of the form
\be
\Lambda^4  \(\frac{\partial}{\Lambda}\)^l  \(\frac{\partial \partial \pi}{\Lambda^3}\)^{n_S} \(\frac{m \partial V}{\Lambda^3}\)^{n_V} \(  \frac{\tilde h}{M} +\frac{m^2 \pi}{\Lambda^3}\)^{n_T} \, ,
\ee
the tree level scattering amplitude will get a contribution
\ba
\Delta A(s) &\sim& g_*^2 \Lambda^4 s^{-2} \( \frac{\sqrt{s}}{\Lambda}\)^l \(\frac{s^{3/2} }{\Lambda^3}\)^{n_S} \(m \frac{s }{\Lambda^3}\)^{n_V} \( \frac{\sqrt{s}}{M} \)^{n_T} \\
&\sim& g_*^2 \frac{m^2}{\Lambda^2}\( \frac{\sqrt{s}}{\Lambda}\)^l  \frac{ m^{n_V+2 n_T} s^{4 - n_V/2-n_T}}{m^2 \Lambda^{6}} \, ,
\ea
and so the contribution to the positivity bound is
\be
\Delta B^{(2,0)}(s= 2m^2) \sim  g_*^2 \frac{m^2}{\Lambda^2}\( \frac{m}{\Lambda}\)^l  \frac{ m^2}{\Lambda^{6}} \, .
\ee
Not unsurprisingly, this is suppressed by at least one power of $m^2/\Lambda^2 = \Lambda/M$ relative to the leading contribution. That is because the $\beta$ terms are assumed to come in at the scale $\Lambda^4$ rather than $\Lambda^3 M$, and it is not to do with the derivative suppression. \\

Let us consider the exchange interactions in which one vertex (left) comes from a leading interaction, and the other vertex (right) comes from the higher derivative interactions. Using the same notation as earlier, we have
\ba
A(s) &\sim& \frac{g_*^2}{s^3} \Lambda^3 M \(\frac{s^{3/2} }{\Lambda^3}\)^{n^L_S+m_S-p} \(\frac{ms }{\Lambda^3}\)^{n^L_V+m_V} \( \frac{\sqrt{s}}{M} \)^{n^L_T+m_T} \( \frac{m^2 \sqrt{s}}{\Lambda^3}\)^p  \nn\\
&\times& \frac{1}{s} \Lambda^4  \( \frac{\sqrt{s}}{\Lambda}\)^l \(\frac{s^{3/2} }{\Lambda^3}\)^{n_S^R+m_S} \(\frac{m s }{\Lambda^3}\)^{n_V^R+m_V} \( \frac{\sqrt{s}}{M} \)^{n_T^R+m_T} \\
&=& g_*^2 \frac{M \Lambda^7}{s^4}\( \frac{\sqrt{s}}{\Lambda}\)^l \(\frac{s^{3/2} }{\Lambda^3}\)^{n_S+2m_S-p} \(\frac{m s }{\Lambda^3}\)^{n_V+2m_V} \( \frac{\sqrt{s}}{M} \)^{n_T+2m_T} \( \frac{m^2 \sqrt{s}}{\Lambda^3}\)^p  \, . \nn
\ea
Again defining $n_P'=n_P+ 2m_P$ and using $\sum_P n_P'=6$ then
\be
A(s) \sim g_*^2 \frac{m^{n_V'+2n_T'+2p} s^{5+l/2-n_T'-n_V'/2-p}}{m^2 \Lambda^{8+l}} \, ,
\ee
which again gives $\Delta B^{(2,0)}(s= 2m^2) \sim  g_*^2 \frac{m^2}{\Lambda^2}\( \frac{m}{\Lambda}\)^l  \frac{ m^2}{\Lambda^{6}} $. \\

In the case where both vertices in the exchange interaction are from higher derivative operators we have
\ba
A(s) &\sim& \frac{g_*^2}{s^3} \Lambda^4  \( \frac{\sqrt{s}}{\Lambda}\)^{l^L} \(\frac{s^{3/2} }{\Lambda^3}\)^{n_S^L+m_S} \(\frac{m s }{\Lambda^3}\)^{n_V^L+m_V} \( \frac{\sqrt{s}}{M} \)^{n_T^L+m_T}  \nn\\
&\times& \frac{1}{s} \Lambda^4  \( \frac{\sqrt{s}}{\Lambda}\)^{l^R} \(\frac{s^{3/2} }{\Lambda^3}\)^{n_S^R+m_S} \(\frac{m s }{\Lambda^3}\)^{n_V^R+m_V} \( \frac{\sqrt{s}}{M} \)^{n_T^R+m_T} \\
&\sim& g_*^2 \frac{ \Lambda^8}{s^4}\( \frac{\sqrt{s}}{\Lambda}\)^l \(\frac{s^{3/2} }{\Lambda^3}\)^{n_S+2m_S} \(\frac{m s }{\Lambda^3}\)^{n_V+2m_V} \( \frac{\sqrt{s}}{M} \)^{n_T+2m_T} \nn \\
&\sim& g_*^2 \frac{m^{n_V'+2n_T'} s^{5+l/2-n_T'-n_V'/2}}{ \Lambda^{10+l}}  \, ,
\ea
with $l= l^L+l^R$. These interactions are typically further $m^2/\Lambda^2$ suppressed and give $\Delta B^{(2,0)}(s= 2m^2) \sim  g_*^2 \frac{m^4}{\Lambda^4}\( \frac{m}{\Lambda}\)^l  \frac{ m^2}{\Lambda^{6}} $ \\

For the next non-trivial positivity bound we have in both of the first two cases
\be
\Delta B^{(4,0)}(s=2m^2) \sim  g_*^2 \( \frac{m}{\Lambda}\)^l  \frac{ 1}{\Lambda^{8}} \, ,
\ee
for which the leading $l=0$ term is exactly at the na\"{i}ve expected order at least for $SSSS$ scattering. Specifically the interaction
\be
\Lambda^4 \( \frac{\partial \partial \pi}{\Lambda^3} \)^4
\ee
will contribute at the desired order (this was discussed in \cite{deRham:2017avq}). For $l=0$ interactions, constraints from one or more vectors and tensors will not contribute as they have fewer derivatives. This forces us to look to $l=2$, for which we recover the $m^2/\Lambda^2$ suppression characteristic of the leading positivity bound
\be
\Delta B^{(4,0)}(s=2m^2) \sim   \frac{m^2}{\Lambda^2 } g_*^2   \frac{ 1}{\Lambda^{8}}  \, , \quad n_V/2+n_T \ge 1 \, , \, l=2 \, .
\ee
This apparent smallness of $B^{(4,0)}(s=2m^2)$ creates its own tension, since again the right hand side of the $B^{(4,0)}$ positivity bound will be finite in the limit $m\rightarrow 0$. However, once again it will be suppressed by $g_*^2$, and given the weak coupling already required by $B^{(2,0)}$, there will in general be no difficulty tuning the higher derivative operators to satisfy it.

\subsection{EFT corrections at one-loop from leading terms}

To estimate the one-loop corrections to the scattering amplitude with $n_S$ scalars, $n_V$ vectors and $n_T$ tensors from the leading interactions, we can use the optical theorem. For each $n_P$ we split it into a left and right part of the unitarity cut $n_P=n_P^L+n_P^R$ and denote by $m_P$ the number of internal polarizations so that
\be
\sum_P n_P^L = \sum_P n_P^R=\sum_P m_P = 2 \, .
\ee
Then from the optical theorem, the form of the one-loop amplitude is
\ba
 A_{\rm one-loop}(s) &\sim
& \sum_{m_P} \(g_*^2 \frac{ m^{n^L_V+m_V+2 n_T^L+2m_T+2 p^L} s^{4 - (n_V^L+m_V)/2-(n_T^L+m_T)-p_L}}{m^2 \Lambda^{6}}\) \times  \(L\to R\)  \nn \\
&\sim&  \sum_{m_P} g_*^4  \frac{s^8 }{m^4\Lambda^{12}} m^{n_V+2m_V+2 n_T+4 m_T+2 p} s^{4 - (n_V+2 m_V)/2-(n_T+2 m_T) -p}\, ,
\ea
with $p=p^L+p^R$ and so the correction to the leading positivity bound is
\be
B^{(2,0)}_{\rm one-loop}(s=2m^2) \sim g_*^4  \frac{m^8 }{\Lambda^{12}}  \, ,
\ee
which is suppressed by $g_*^2 m^6/\Lambda^6$ relative to the tree level contribution. This was to be expected as the effective loop counting parameter at energy $E$ is $g_*^2 E^6/\Lambda^6$. As discussed at length in section \ref{Mixing}, the fact that the loop contribution is suppressed is at the origin of the mixing of orders, by which the one loop imaginary part of the amplitude constrains the size of the tree level real part via analyticity. Loop corrections from higher derivative interactions will automatically be additionally suppressed.

\section{Entering the Vainshtein region in Weakly Coupled massive gravity}
\label{sec:EnteringVain}

The discussion of the Vainshtein region for massive gravity, closely parallels that for the Galileon with $K_{\mu\nu}$ playing the role of the order parameter $\partial_{\mu} \partial_{\mu} \pi/\Lambda^3$.
Reiterating the structure of the EFT for massive gravity
 \be
 \label{Kcorrections1}
{\cal L} = \frac{1}{g_*^2}\left( \Lambda^4 L_0 + g_*^2 L_1 + \dots \right)
\ee
\be
\label{Kcorrections}
\Lambda^4 L_0 =  \left[  \frac{M^2}{2} R  - \Lambda^3 M \sum_{n} \alpha_n {\cal E} {\cal E} g^{4-n} K^n \right] + \Lambda^4 \sum \beta_{p,q,r} \( \frac{\nabla}{\Lambda}\)^p K_{\mu\nu}^{q} \( \frac{R_{\mu\nu \rho \sigma}}{\Lambda^2} \)^r \, ,
\ee
the boundary of the Vainshtein region is defined as the point at which the locally inertial values of the Riemann curvature, $R_{\mu\nu \rho \sigma}$, become comparable to $m^2$. For a point source of mass $M_0$, this is when $M_0/(\mpl ^2r^3) \sim m^2$, which is the same Vainshtein radius for the Galileon. At this point, the equations of motion for the leading terms will imply the locally inertial value of $K_{\mu\nu} \sim 1$. The leading terms will then give order unity renormalizations to the kinetic terms for the fluctuations.
This is consistent with the expectations of the decoupling limit. At this point the $\beta$ term corrections to the tree level effective action are suppressed by at least
\be
\frac{L_{\text{higher derivative}}}{L_{\rm leading}} \sim \frac{\Lambda^4}{m^2 M^2} = \frac{\Lambda}{M} =  \frac{m^2}{\Lambda^2} \sim 10^{-30} \ll 1 \, .
\ee
This allows us to enter some way into the Vainshtein region without the EFT expansion losing complete control.  From the point of view of the low energy expansion, the regime of convergence is $K \ll 1$, $\nabla \ll \Lambda$ and $R \ll \Lambda^2$. However we can again suppose that there is a resummation of the EFT corrections so that around a background the cutoff is effectively raised. \\

As in the previous example, this can be achieved by imagining that the mass of the heavy fields scales with a positive power of $K$, for instance
\be
{\cal L}_H \sim   \frac{1}{g_*^2} \left( -\frac{1}{2}\partial_{\mu} H \partial_{\nu} H g^{\mu\nu} - \frac{1}{2}\Lambda^2 \( 1+ {\rm Tr}[K]\)  H^2  + \lambda \Lambda^3 H {\rm Tr}[K] \dots \right)\,,
\ee
which on integrating out at tree level will generate higher derivative interactions automatically suppressed by $(\Lambda \sqrt{(1+\text{Tr}[K])})^{-1} \partial  $. In practice, we may need a large number of heavy $H$ fields (possibly even infinitely many) of arbitrary spins to provide an explicit UV completion, but we can easily imagine that the masses of all such states scale in the appropriate way.\\

Alternatively, we can imagine that all the heavy fields couple to the effective metric $g^{\rm eff}_{\mu\nu}= g_{\mu\nu}+ \alpha K_{\mu\nu}$ and have masses at the scale $\Lambda$. Then integrating out such heavy states at loop level, the contribution to the Lagrangian will take the form
\be
 \Delta L_1 = \Lambda^4 \sqrt{{\rm Det}(\delta\mupn+\alpha K\mupn)}\  L_H\left[\frac{\nabla_{\mu}^{\rm eff}}{\Lambda}, R^{\mu}{}_{\nu \rho \sigma}[g_{\rm eff}] \right] \, ,
\ee
where $\nabla_{\mu}^{\rm eff}$ is the covariant derivative with respect to the effective metric and the notation signifies that all contractions are performed with respect to the effective metric. Crucially both these special structures are completely consist with the perturbative form (\ref{Kcorrections1},\ref{Kcorrections}), valid for $K \ll 1$.
\\

In the second example, as we go deep into the Vainshtein region, so that $g_{\rm eff} \sim \alpha K \eta_{\mu\nu}$ the derivative and curvature corrections effectively scale as $\nabla^{\rm eff} \sim \frac{1}{\sqrt{K}} \partial $ and $R \sim \frac{1}{K} $ (consider them evaluated in a locally inertial frame), which is equivalent to saying that the effective scale of derivative interactions is now $\sqrt{K} \Lambda$.
This assumption is a tuning, but one that is technically natural from the perspective of the loops of the heavy fields if it is assumed that all the heavy states responsible for spontaneously breaking diffeomorphisms couple to the same effective metric. In this sense the special resummed structure necessary to exhibit the Vainshtein mechanism can itself be potentially technically natural, at least from the perspective of heavy loops, for weakly coupled UV completions\footnote{Loops of light fields are always best computed in dimensional regularization to avoid tracking misleading cutoff dependence which may not be truly reflective of the heavy physics \cite{Burgess:1992gx}.}. It is also apparent from these ansatze that the derivative corrections for all the helicity components of the massive graviton will be appropriately rescaled, and so the resummation is not something confined to the scalar modes.

\paragraph{Recovering GR at small distances:}

In these conjectured weakly coupled UV completions of massive Galileons and massive gravity, the onset of the breakdown of the EFT occurs at distance scales smaller than the Vainshtein radius. This loss of predictivity means that we cannot use the current formulations of massive gravity to make predictions---for example for table top tests of gravity. However, it does not mean that the Vainshtein mechanism is not active in whatever appropriately UV completes the theory. The loss of predictivity comes from the breakdown of the derivative expansion, which signals the existence of new massive states, and not from strong coupling (i.e. not the breakdown of the loop expansion). For both the Galileon and massive gravity EFTs, it is essential to keep in mind that
\begin{enumerate}
  \item The virtue of the Vainshtein mechanism is to decouple the strongly self-interacting sector. This means that while loss of predictivity due to new states or strong coupling occurs, this occurs not for the standard gravitational sector nor the Standard Model, but for a sector which decouples and is already suppressed by several (9 in the most pessimistic estimates) orders of magnitude upon entering the region where the EFT breaks down.
  \item Whatever occurs when the EFT breaks down, it is highly unlikely that  the new physical states or strong coupling would suddenly start (a) reverting the decoupling that was already well underway and (b) involve a long-range force with physical observables.
  \item Loss of predictivity from a given energy scale onwards can by no means be used to rule in or out the validity of an EFT at lower energy scales.
  \item All these considerations follow approach \ref{appro3} as an original hypothesis, which may or may not be the natural embedding for these types of theories.
\end{enumerate}

\vskip 10pt
\acknowledgments
We would like to thank Brando Bellazzini, David Pirtskhalava, Riccardo Rattazzi, Francesco Riva, Javier Serra, Francesco Sgarlata, Michael Trott and Shuang-Yong Zhou for useful comments. The work of CdR and AJT is supported by STFC grant ST/P000762/1. CdR thanks the Royal Society for support at ICL through a Wolfson Research Merit Award. CdR is also supported in part by the European Union's Horizon 2020 Research Council grant 724659 MassiveCosmo ERC-2016-COG and in part by a Simons Foundation award ID 555326 under the Simons Foundation’s Origins of the Universe initiative, `{\it Cosmology Beyond Einstein's Theory}'. SM is funded by the Imperial College President's Fellowship. AJT thanks the Royal Society for support at ICL through a Wolfson Research Merit Award.

\bibliographystyle{JHEP}
\bibliography{references}

\end{document}